\documentclass[showpacs,aps,prb,lettersize]{revtex4}
\usepackage{amsmath}
\usepackage{graphicx}

\pacs{71.10.Fd, 05.30.Fk, 71.10.-w}

\begin{document}

\title{Non-Perturbative Treatment of Charge and Spin Fluctuations in the
Two-Dimensional Extended Hubbard Model: Extended Two-Particle
Self-Consistent Approach}
\author{B. Davoudi$^{1,2}$ and A.-M.S. Tremblay$^1$}

\begin{abstract}
We study the spin and charge fluctuations of the extended Hubbard model
(EHM) with on-site interaction $U$ and first neighbor interaction $V$ on the
two-dimensional square lattice in the weak to intermediate coupling regime.
We propose an extension of the two-particle self-consistent (ETPSC)
approximation that includes the effect of functional derivatives of the pair
correlation functions on irreducible spin and charge vertices. These
functional derivatives were ignored in our previous work. We evaluate them
assuming particle-hole symmetry. The resulting theory satisfies
conservations laws and the Mermin-Wagner theorem. Our current results are in
much better agreement with benchmark Quantum Monte-Carlo (QMC) results. As a
function of $U$ and $V,$ we can determine the crossover temperatures towards
renormalized classical regimes where either spin or charge fluctuations
dominate. The dominant wave vector is self-determined by the approach.
\end{abstract}

\maketitle

\renewcommand{\thefigure}{\arabic{figure}} 

\address{$^1$ D\'epartment de Physique and Centre de Recherche en Physiqiue du Solide, Universit\'e de Sherbrooke, Sherbrooke, Qu\'ebec, Canada J1K 2R1\\
$^2$Institute for Studies in Theoretical Physics and Mathematics, Tehran 19395-5531, Iran}

\section{Introduction}

\label{sec1}

The single-band Hubbard model, with on-site interaction $U$ and
nearest-neighbor hopping $t,$ is the simplest model of correlated electrons.
It has been extensively studied, for example, in the context of
high-temperature superconductors. Charge ordering phenomena in strongly
correlated systems are also quite common, for example in manganites,
vanadates, cobaltates and various organic conductors. \cite{Imada:2005,Oles}
In these cases, and even in high temperature superconductors where screening
is not perfect, it is useful to investigate the Extended Hubbard model (EHM)
that includes the effect of nearest-neighbor Coulomb repulsion $V$. In that
model, charge and spin fluctuations compete over much of the phase diagram.

Even at weak to intermediate coupling, $\left( U\text{ and }V\text{ less
than the bandwidth }W=8t\text{ in }d=2\right) $ very few many-body methods
can treat the Hubbard model while adequately taking into account the
Mermin-Wagner theorem. The Fluctuation-Exchange Approximation (FLEX) \cite%
{Bickers:1989} is such a method. The Two-Particle Self-Consistent approach
(TPSC) \cite{Vilk1,Vilk2,Mahan,Hankevych} is an alternate non-perturbative
approach that compares much better with benchmark Quantum Monte Carlo (QMC)
simulations. In addition, contrary to FLEX, it reproduces the pseudogap
phenomenon observed in QMC. \cite{Moukouri,AllenKyung}

In this paper, we present an extension of the TPSC approach that allows one
to treat the EHM. The accuracy of the new approach is checked by comparing
with QMC simulations. \cite{Zhang} The competition between charge and spin
fluctuations in parameter space is also explored for two fillings. Whenever
possible we compare with results of other approaches available in the
literature. Studies for the specific parameters appropriate for given
compounds will be the subject of separate papers. The purpose of the present
one is to give a detailed description of the theoretical approach, which is
an extension of the TPSC approach.

Briefly speaking, the TPSC method was derived from a functional approach by
approximating the vertex functions as local functions of the imaginary time
and space (in other words irreducible vertices independent of frequency and
wave vector). This leads to irreducible spin and charge vertices that
contain two unknown constants, namely the on-site pair correlation function
and its functional derivative, which can be self-consistently determined
from two sum-rules on the structure functions. Unfortunately the number of
unknowns becomes more than the number of available sum-rules in the EHM, so
one has to use another approach. In our previous work, \cite{Bahman} we
simply neglected the contribution of functional derivatives of the pair
correlation functions. We obtained good agreement with QMC results for small
values of $V$ when spin fluctuations are still dominant, but this approach
did not work so well in the opposite limit when charge fluctuations becomes
dominant. This is mainly due to ignoring the functional derivative of the
pair correlation functions. In this paper, we demonstrate the importance of
these functional derivative terms, explain the way they enter in the
formalism and propose a way to use particle-hole symmetry to evaluate them.

The EHM has been studied within different approaches and approximations. We
can only give a sample of the literature. Many papers focus on the
quarter-filled case on the square lattice or on ladders \cite%
{Seo:1998,Pietig:1999, Vojta:1999,
Hellberg:2001,Vojta:2001,Hoang:2002,Calandra:2002,Imada:2005}. The
competition between charge and spin orders has also been studied in other
dimension: one- \cite{Fourcade, Bosch:1988, Aichhorn:2004}, three- \cite%
{Barktowiak:1995} and higher dimensions \cite{VanDongen:1994} at various
fillings. In two dimensions, there are several studies at strong coupling
\cite{Avella:2004,Avella:2006,Vojta:2002} or on lattices different from the
square lattice. \cite{Kuroki:2006,Seo:2006} The combined effect of charge
fluctuations in addition to the usual spin fluctuation in favouring one type
or another of unconventional superconductivity has also been studied using
this model \cite{Onari, Kishine:1995, Callaway:1990,Scalapino,Esirgen}.

On the two-dimensional square lattice, QMC studies \cite{Zhang} will be
particularly relevant as mentioned above. We have not found any systematic
comparison between FLEX approximation for the EHM \cite{Onari} and QMC. The
mean field (MF) approximation \cite{Yan:2006} and the renormalization group
(RG) approach \cite{Murakami}, as well as exact diagonalization \cite%
{Ohta:1994}, perturbation theory about ordered states \cite{Yan:1993} and
trial correlated wave functions \cite{Oles} have been applied to this
problem. One of the results that we will find at half-filling is a crossover
line at $U\approx 4V$ from dominant antiferromagnetic fluctuations to
dominant checkerboard charge fluctuations. This is in agreement with several
studies \cite{Zhang,Onari,Oles,Ohta:1994,Yan:1993,Bari:1971} as we will
discuss. Our results hold at the crossover temperature. Most other
approaches focus on the ground state. The result $U=zV,$ where $z$ is the
number of nearest-neighbors, seems to hold quite generally in various
dimensions. \cite{Oles,Fourcade,Micnas} Some MF results \cite{Murakami} give
the possibility of a coexistence region between charge and spin density
waves. The Pauli principle prevents charge and spin fluctuations to increase
at the same time in the paramagnetic regime at finite temperature. This is
satisfied within our approach.

In the following we present the formalism, which first focuses on finding a
set of closed equations for the irreducible vertices. We explain how the
functional derivative of the pair correlation functions enter in the
evaluation of irreducible vertices. In the second part of the theory
section, we explain how particle-hole symmetry can be used to estimate the
functional derivatives and we give the details of the new resulting Extended
TPSC (ETPSC) approach. In Sec. \ref{sec4}, we present numerical results and
compare them with available QMC results to gauge the accuracy of the
approach. The last part of that section presents a systematic study on the
square lattice of the crossover to the renormalized classical regimes for
charge and spin fluctuations for two densities, $n=1$ and $n=0.75$. Finally,
in the concluding section we discuss the possibility to use the same type of
approximation for other systems and point towards possible future
developments and improvements.

\section{Theory}

\subsection{Some exact results, factorization and approximation for the
functional derivatives}

\label{sec2} The extended Hubbard Hamiltonian is defined by,  
\begin{eqnarray}
H &=&-t\sum_{\left\langle \mathbf{ij}\right\rangle \sigma }(c_{\mathbf{i}%
\sigma }^{\dagger }c_{\mathbf{j}\sigma }+c_{\mathbf{j}\sigma }^{\dagger }c_{%
\mathbf{i}\sigma })+U\sum_{\mathbf{i}}n_{\mathbf{i}\uparrow }n_{\mathbf{i}%
\downarrow } \\
&&+V\sum_{\left\langle \mathbf{ij}\right\rangle \sigma \sigma ^{\prime }}n_{%
\mathbf{i}\sigma }n_{\mathbf{j}\sigma ^{\prime }}-\mu \sum_{\mathbf{i}}n_{i}
\end{eqnarray}%
where $c_{\mathbf{i}\sigma }$ ($c_{\mathbf{i}\sigma }^{\dagger }$) are
annihilation (creation) operators for an electron with spin $\sigma $ in the
Wannier orbital at site $i$, $n_{\mathbf{i}\sigma }=c_{\mathbf{i}\sigma
}^{\dagger }c_{\mathbf{i}\sigma }$ is the density operator for electrons of
spin $\sigma $, $t$ is the hopping matrix element and $\mu $ is the chemical
potential. The terms proportional to $U$ and $V$ represent respectively the
on-site and nearest-neighbor interactions. The latter $\left( V\right) $
vanishes for the usual Hubbard model.

To generalize the TPSC approach, we use the same formalism as in Ref. 
\onlinecite{Allen:2003}%
. The Green function is defined by 
\begin{eqnarray}
G_{\sigma }(1,2) &=&-\frac{\delta \ln Z[\phi _{\sigma }]}{\delta \phi
_{\sigma }(2,1)}=-\left\langle T_{\tau }c_{1\sigma }(\tau _{1})c_{2\sigma
}^{\dagger }(\tau _{2})\right\rangle  \\
&=&-\left\langle T_{\tau }c_{\sigma }(1)c_{\sigma }^{\dagger
}(2)\right\rangle 
\end{eqnarray}%
where the brackets $\left\langle {}\right\rangle $ represents a thermal
average in the canonical ensemble, $T_{\tau }$ is the time-ordering
operator, and $\tau $ is the imaginary time. The space and imaginary time
coordinates $\left( \mathbf{r}_{1},\tau _{1}\right) $ are abbreviated as $%
\left( 1\right) .$ The generating functional $Z[\phi _{\sigma }]$ is given
by 
\begin{equation}
Z[\phi _{\sigma }]=\mathrm{Tr}\left[ e^{-\beta H}T_{\tau }S(\beta )\right] 
\end{equation}%
and $S(\beta )$ is defined as follows 
\begin{equation}
\ln S(\beta )=-\sum_{\mathbf{i},\mathbf{j},\sigma }\int_{0}^{\beta }d\tau
d\tau ^{\prime }c_{\mathbf{i}\sigma }^{\dagger }(\tau )c_{\mathbf{j}\sigma
}(\tau ^{\prime })\phi _{\sigma }(\mathbf{i},\mathbf{j},\tau ,\tau ^{\prime
}).
\end{equation}

The equation of motion for the Green function takes the form of the Dyson
equation \cite{BaymKadanoff:1962, Vilk2} 
\begin{equation}
G_{\sigma }^{-1}(1,2)=G_{0}^{-1}(1,2)-\phi _{\sigma }(1,2)-\Sigma _{\sigma
}(1,2).  \label{Dyson}
\end{equation}%
The expression for $\Sigma G$ comes from the commutator with the interaction
part of the Hamiltonian so that the self-energy in the above equation is
given by%
\begin{widetext}%
\begin{equation}
\Sigma _{\sigma }(1,2)=-U\left\langle T_{\tau }c_{\tilde{\sigma}}^{\dagger
}(1)c_{\tilde{\sigma}}(1)c_{\sigma }(1)c_{\sigma }^{\dagger }(\bar{3}%
)\right\rangle G_{\sigma }^{-1}(\bar{3},2)-V\sum_{\sigma ^{\prime
},a}\left\langle T_{\tau }c_{\sigma ^{\prime }}^{\dagger }(1+a)c_{\sigma
^{\prime }}(1+a)c_{\sigma }(1)c_{\sigma }^{\dagger }(\bar{3}),\right\rangle
G_{\sigma }^{-1}(\bar{3},2)
\end{equation}%
where $\tilde{\sigma}=-\sigma $ and the summation on $a$ runs over the next
neighbor sites of the site $1$ (at the same imaginary time).

The first approximation is inspired by TPSC. We factorize the two-body
density matrix operators as follows, defining numbers with overbars, such as 
$\bar{3},$ to mean a sum over the corresponding spatial indices and
integration over the corresponding imaginary time: 
\begin{align}
\Sigma _{\sigma }(1,2)& \cong UG_{\tilde{\sigma}}(1,1^{+})G_{\sigma }(1,\bar{%
3})G_{\sigma }^{-1}(\bar{3},2)g_{\sigma \tilde{\sigma}}(1,1)+V\sum_{\sigma
^{\prime },a}G_{\sigma ^{\prime }}(1+a,1+a^{+})G_{\sigma }(1,\bar{3}%
)G_{\sigma }^{-1}(\bar{3},2)g_{\sigma \sigma ^{\prime }}(1,1+a)  \notag \\
& =U\delta (1,2)G_{\tilde{\sigma}}(1,1^{+})g_{\sigma \tilde{\sigma}%
}(1,1)+V\delta (1,2)\sum_{\sigma ^{\prime },a}G_{\sigma ^{\prime
}}(1+a,1+a^{+})g_{\sigma \sigma ^{\prime }}(1,1+a).  \label{Sigma+}
\end{align}%
\end{widetext}%
If we set $g_{\sigma \sigma ^{\prime }}(i,j)=1$ in the above expression, one
recovers the Hartree-Fock approximation for the $U$ term, and the Hartree
approximation for the $V$ term. The Fock term in the near neighbor
interaction $V$ leads to a renormalization of the band parameters and is not
as essential as in the on-site term, as discussed in our previous paper.\cite%
{Bahman} The above expression for the self-energy goes beyond the
Hartree-Fock approximation because of the presence of the pair correlation
functions $g_{\sigma \sigma ^{\prime }}(i,j)$ defined by 
\begin{equation}
g_{\sigma \sigma ^{\prime }}(1,2)\equiv \frac{\left\langle n_{\sigma }\left(
1\right) n_{\sigma ^{\prime }}\left( 2\right) \right\rangle -\delta
(1,2)\delta _{\sigma ,\sigma ^{\prime }}\left\langle n_{\sigma }\left(
1\right) \right\rangle }{\left\langle n_{\sigma }\left( 1\right)
\right\rangle \left\langle n_{\sigma ^{\prime }}\left( 2\right)
\right\rangle }.  \label{Def_g_sig_sigp}
\end{equation}%
This quantity is related to the probability of finding one electron with
spin $\sigma ^{\prime }$ on site $j$ when another electron with spin $\sigma 
$ is held on site $i$. The presence of the pair correlation function is
analog to the local field correction in the Singwi approach to the electron
gas. \cite{Singwi} Thanks to this correction factor, we recover the exact
result that the combination $\Sigma _{\sigma }(1,\bar{3})G_{\sigma }(\bar{3}%
,2)$ is equal to the potential energy when point $2$ becomes $1^{+}$ (where
the superscript $+$ refers to the imaginary time and serves to order the
corresponding field operator to the left of operators at point $1$). For
this special case then, $g_{\sigma \sigma ^{\prime }}(1,1+a)$ contains the
information about the Fock contribution as well.

We need the response functions obtained from the first functional derivative
of the Green function with respect to the external field. Taking the
functional derivative of the identity $G_{\sigma }(1,\bar{3})G_{\sigma
^{\prime }}^{-1}(\bar{3},2)=\delta _{\sigma \sigma ^{\prime }}\delta (1,2)$
and using Dyson's equation Eq.~(\ref{Dyson}), we find%
\begin{widetext}%
\begin{align}
\chi _{\sigma \sigma ^{\prime }}(1,2;3,3)& \equiv -\frac{\delta G_{\sigma
}(1,2)}{\delta \phi _{\sigma ^{\prime }}(3,3)}=G_{\sigma }(1,\bar{4})\frac{%
\delta G_{\sigma }^{-1}(\bar{4},\bar{5})}{\delta \phi _{\sigma ^{\prime
}}(3,3)}G_{\sigma }(\bar{5},2)  \notag  \label{chi} \\
& =-\delta _{\sigma \sigma ^{\prime }}G_{\sigma }(1,3)G_{\sigma
}(3,2)-\sum_{\sigma ^{\prime \prime }}G_{\sigma }(1,\bar{4})\frac{\delta
\Sigma _{\sigma }(\bar{4},\bar{5})}{\delta G_{\sigma ^{\prime \prime }}(\bar{%
6},\bar{7})}\frac{\delta G_{\sigma ^{\prime \prime }}(\bar{6},\bar{7})}{%
\delta \phi _{\sigma ^{\prime }}(3,3)}G_{\sigma }(\bar{5},2).
\end{align}%
The irreducible vertex is the first functional derivative of the self-energy
respect to the Green function, as can be seen from the last equation. It can
be obtained from our equation for the self-energy Eq.~(\ref{Sigma+}):%
\begin{align}
\frac{\delta \Sigma _{\sigma }(4,5)}{\delta G_{\sigma ^{\prime \prime }}(6,7)%
}& =U\delta _{\tilde{\sigma}\sigma ^{\prime \prime }}\delta (4,5)\delta
(4,6)\delta (5,7)g_{\sigma \tilde{\sigma}}(4,4)+V\sum_{a}\delta (4,5)\delta
(4+a,6)\delta (5+a,7)g_{\sigma \sigma ^{\prime \prime }}(4,4+a)  \notag
\label{dsdg} \\
& +U\delta (4,5)G_{\tilde{\sigma}}(4,4^{+})\frac{\delta g_{\sigma \tilde{%
\sigma}}(4,4)}{\delta G_{\sigma ^{\prime \prime }}(6,7)}+V\delta
(4,5)\sum_{\sigma ^{\prime \prime \prime },a}G_{\sigma ^{\prime \prime
\prime }}(4,4^{+})\frac{\delta g_{\sigma \sigma ^{\prime \prime \prime
}}(4,4+a)}{\delta G_{\sigma ^{\prime \prime }}(6,7)}.
\end{align}

\end{widetext}%
In the standard Hubbard model with $V=0$, there is only one unknown
functional derivative of the pair correlation function with respect to the
Green function. Since it enters only the charge response function , it can
be determined by making the local approximation, enforcing the Pauli
principle and requiring that the fluctuation-dissipation theorem relating
charge susceptibility to pair correlation function be satisfied. \cite{Vilk2}
In the present case, there are not enough sum rules to determine the
additional functional derivatives of the pair correlation functions with
respect to the Green function.

In our previous paper on the extended Hubbard model, \cite{Bahman} we
neglected these additional unknown functional derivatives. Here we present
an approximation for these functional derivatives that improves our previous
results, as can be judged by comparisons with QMC calculations. We would
like to replace the unknown terms by functional derivatives with respect to
the density as $\frac{\delta g_{\sigma \sigma ^{\prime }}(1,2)}{\delta
G_{\sigma ^{\prime \prime }}(3,4)}\approx \frac{\delta g_{\sigma \sigma
^{\prime }}(1,2)}{\delta n_{\sigma ^{\prime \prime }}(3)}\delta (3,4)$.
Unfortunately we cannot evaluate this type of functional derivative either,
unless point $3\ $coincides with either $2$ or $1$, so we approximate the
previous equation by 
\begin{equation}
\frac{\delta g_{\sigma \sigma ^{\prime }}(1,2)}{\delta n_{\sigma ^{\prime
\prime }}(3)}\approx \frac{\delta g_{\sigma \sigma ^{\prime }}(1,2)}{\delta
n_{\sigma ^{\prime \prime }}(1)}\delta (1,3)+\frac{\delta g_{\sigma \sigma
^{\prime }}(1,2)}{\delta n_{\sigma ^{\prime \prime }}(2)}\delta (2,3).
\label{Functional_of_density}
\end{equation}%
In other words, we ignore the terms where point $3$ gets further away from
either $1$ or $2$. We will show that the last equation is exact at
half-filling (see Eq.(\ref{Vanishing_fd})) and that it can be evaluated . We
will also show in the next section by explicit comparisons with QMC
calculations that its extension away from half-filling is a reasonable
approximation. The last two approximations lead to the local approximation of TPSC as
a special case.

Inserting in the susceptibility Eq.~(\ref{chi}) the irreducible vertex
obtained from Eq.~(\ref{dsdg}) and our approximation for the functional
derivatives, we find%
\begin{widetext}%

\begin{align}
\chi _{\sigma \sigma ^{\prime }}(1;2)=& -\delta _{\sigma \sigma ^{\prime
}}G_{\sigma }(1,2)G_{\sigma }(2,1)  \notag \\
& -UG_{\sigma }(1,\bar{3})G_{\sigma }(\bar{3},1)g_{\sigma \tilde{\sigma}}(%
\bar{3},\bar{3})\frac{\delta G_{\tilde{\sigma}}(\bar{3},\bar{3}^{+})}{\delta
\phi _{\sigma ^{\prime }}(2,2)}  \notag \\
& -VG_{\sigma }(1,\bar{3})G_{\sigma }(\bar{3},1)\sum_{\sigma ^{\prime \prime
},a}g_{\sigma \sigma ^{\prime \prime }}(\bar{3},\bar{3}+a)\frac{\delta
G_{\sigma ^{\prime \prime }}(\bar{3}+a,\bar{3}+a^{+})}{\delta \phi _{\sigma
^{\prime }}(2,2)}  \notag \\
& -UG_{\sigma }(1,\bar{3})G_{\sigma }(\bar{3},1)n_{\tilde{\sigma}}(\bar{3}%
)\sum_{\sigma ^{\prime \prime }}\frac{\delta g_{\sigma \tilde{\sigma}}(\bar{3%
},\bar{3})}{\delta n_{\sigma ^{\prime \prime }}(\bar{3})}\frac{\delta
G_{\sigma ^{\prime \prime }}(\bar{3},\bar{3}^{+})}{\delta \phi _{\sigma
^{\prime }}(2,2)}  \notag \\
& -VG_{\sigma }(1,\bar{3})G_{\sigma }(\bar{3},1)\sum_{\sigma ^{\prime \prime
}\sigma ^{\prime \prime \prime }a}n_{\sigma ^{\prime \prime \prime }}(\bar{3}%
)\left[ \frac{\delta g_{\sigma \sigma ^{\prime \prime \prime }}(\bar{3},\bar{%
3}+a)}{\delta n_{\sigma ^{\prime \prime }}(\bar{3})}\frac{\delta G_{\sigma
^{\prime \prime }}(\bar{3},\bar{3}^{+})}{\delta \phi _{\sigma ^{\prime
}}(2,2)}+\frac{\delta g_{\sigma \sigma ^{\prime \prime \prime }}(\bar{3},%
\bar{3}+a)}{\delta n_{\sigma ^{\prime \prime }}(\bar{3}+a)}\frac{\delta
G_{\sigma ^{\prime \prime }}(\bar{3}+a,\bar{3}+a^{+})}{\delta \phi _{\sigma
^{\prime }}(2,2)}\right] .
\end{align}%
For the repulsive case, we expect that spin and charge components of the
response functions will be dominant. Thus we need the dynamic correlation
functions of the density $n=n_{\uparrow }+n_{\downarrow }$ and magnetization
$m=n_{\uparrow }-n_{\downarrow }$ respectively. They are given by
\begin{align}
\chi _{cc,ss}(1;2)=& 2\left[ \chi _{\sigma \sigma }(1;2)\pm \chi _{\sigma
\tilde{\sigma}}(1;2)\right]   \notag \\
=& -2G_{\sigma }(1,2)G_{\sigma }(2,1)  \notag \\
& -2UG_{\sigma }(1,\bar{3})G_{\sigma }(\bar{3},1)g_{\sigma \tilde{\sigma}}(%
\bar{3},\bar{3})\left[ \frac{\delta G_{\sigma }(\bar{3},\bar{3}^{+})}{\delta
\phi _{\tilde{\sigma}}(2,2)}\pm \frac{\delta G_{\sigma }(\bar{3},\bar{3}^{+})%
}{\delta \phi _{\sigma }(2,2)}\right]   \notag \\
& -4VG_{\sigma }(1,\bar{3})G_{\sigma }(\bar{3},1)\sum_{a}g_{cc,ss}(\bar{3},%
\bar{3}+a)\left[ \frac{\delta G_{\sigma }(\bar{3}+a,\bar{3}+a^{+})}{\delta
\phi _{\tilde{\sigma}}(2,2)}\pm \frac{\delta G_{\sigma }(\bar{3}+a,\bar{3}%
+a^{+})}{\delta \phi _{\sigma }(2,2)}\right]   \notag \\
& -2nUG_{\sigma }(1,\bar{3})G_{\sigma }(\bar{3},1)\left\{ \frac{\delta
g_{\sigma \tilde{\sigma}}(\bar{3},\bar{3})}{\delta n(\bar{3})},\frac{\delta
g_{\sigma \tilde{\sigma}}(\bar{3},\bar{3})}{\delta m(\bar{3})}\right\} \left[
\frac{\delta G_{\sigma }(\bar{3},\bar{3}^{+})}{\delta \phi _{\tilde{\sigma}%
}(2,2)}\pm \frac{\delta G_{\sigma }(\bar{3},\bar{3}^{+})}{\delta \phi
_{\sigma }(2,2)}\right]   \notag \\
& -4nVG_{\sigma }(1,\bar{3})G_{\sigma }(\bar{3},1)\sum_{a}\left\{ \frac{%
\delta g_{s\sigma }(\bar{3},\bar{3}+a)}{\delta n(\bar{3})},\frac{\delta
g_{s\sigma }(\bar{3},\bar{3}+a)}{\delta m(\bar{3})}\right\} \left[ \frac{%
\delta G_{\sigma }(\bar{3},\bar{3}^{+})}{\delta \phi _{\tilde{\sigma}}(2,2)}%
\pm \frac{\delta G_{\sigma }(\bar{3},\bar{3}^{+})}{\delta \phi _{\sigma
}(2,2)}\right]   \notag \\
& -4nVG_{\sigma }(1,\bar{3})G_{\sigma }(\bar{3},1)\sum_{a}\left\{ \frac{%
\delta g_{s\sigma }(\bar{3},\bar{3}+a)}{\delta n(\bar{3}+a)},\frac{\delta
g_{s\sigma }(\bar{3},\bar{3}+a)}{\delta m(\bar{3+a})}\right\} \left[ \frac{%
\delta G_{\sigma }(\bar{3}+a,\bar{3}+a^{+})}{\delta \phi _{\tilde{\sigma}%
}(2,2)}\pm \frac{\delta G_{\sigma }(\bar{3}+a,\bar{3}+a^{+})}{\delta \phi
_{\sigma }(2,2)}\right]
\end{align}%
\end{widetext}%
where the first term $A$ in the curly brackets $\left\{ A,B\right\} $ is
associated with the upper $+$ sign, and the second term $B$ is associated
with the $-$ sign. Also, in the above equation,
\begin{equation*}
g_{cc,ss}=(g_{\sigma \sigma }\pm g_{\sigma \tilde{\sigma}})/2
\end{equation*}%
are the charge and spin correlation functions, while
\begin{equation*}
g_{s\sigma }=(g_{\sigma \sigma }+g_{\sigma \tilde{\sigma}})/2
\end{equation*}%
is the symmetric combination of the spin-resolved pair correlation
functions.  Indeed, the definition of $g_{cc}$ in partially spin
polarized system is different from the above equation and that is the reason
for defining $g_{s\sigma }$. In fact $g_{cc}$ is symmetric function under a
flip of the spins of the electrons but not $g_{s\sigma }$. This is an
important issue when we deal with the functional derivative of the pair
correlation functions with respect to the magnetization. The functional
derivative terms with respect to magnetization in the fourth and the last
term are zero . This is due to the fact that $g_{\sigma \tilde{\sigma}}(1,1)$
and $g_{s\sigma }(1,2)$ are even functions under exchange of $\sigma
\leftrightarrow \tilde{\sigma}$ on site $1$ and on site $2$ respectively.
The functional derivative with respect to density in the fifth and last term are equal
(this can also be shown by following the procedure that we explain in the
following subsection). Now we can factorize
the charge and spin response functions from the above
equation. The final form of the response functions in Fourier-Matsubara
space are given by
\begin{equation}
\chi _{cc}(\mathbf{q},\omega _{n})=\frac{\chi ^{0}(\mathbf{q},\omega _{n})}{%
1+\frac{\chi ^{0}(\mathbf{q},\omega _{n})}{2}U_{cc}(\mathbf{q})},
\label{chi_cc}
\end{equation}%
\begin{equation}
\chi _{ss}(\mathbf{q},\omega _{n})=\frac{\chi ^{0}(\mathbf{q},\omega _{n})}{%
1-\frac{\chi ^{0}(\mathbf{q},\omega _{n}))}{2}U_{ss}(\mathbf{q})},
\label{chi_ss}
\end{equation}%
where %
\begin{widetext}
\begin{align}
U_{cc}(\mathbf{q})& =U\left( g_{\sigma \tilde{\sigma}}(1,1)+n\frac{\delta
g_{\sigma \tilde{\sigma}}(1,1)}{\delta n(1)}\right) +4V\left(
g_{cc}(1,1+a)\gamma (\mathbf{q})+n\frac{\delta g_{s\sigma }(1,1+a)}{\delta
n(1)}(2+\gamma (\mathbf{q}))\right) ,  \label{Ucc} \\
U_{ss}(\mathbf{q})& =Ug_{\sigma \tilde{\sigma}}(1,1)-4V\left(
g_{ss}(1,1+a)\gamma (\mathbf{q})+2n\frac{\delta g_{s\sigma }(1,1+a)}{\delta
m(1)}\right) .  \label{Uss}
\end{align}%
\end{widetext}%
In this equation, $\gamma (\mathbf{q})$ is given by $\gamma (\mathbf{q}%
)=\sum_{\alpha }\cos (q_{\alpha }a)$, $\alpha =1..D$, $D$ being the
dimension of the system, $\omega _{n}=2\pi nT$ is the Matsubara frequency
and $\chi ^{0}(\mathbf{q},\omega _{n})$ is the free response function
\begin{equation}
\chi ^{0}(\mathbf{q},\omega _{n})=\int_{BZ}\frac{d\mathbf{p}}{\nu }\frac{%
f^{0}(\mathbf{p}+\frac{\mathbf{q}}{2})-f^{0}(\mathbf{p}-\frac{\mathbf{q}}{2})%
}{i\omega _{n}-\epsilon _{\mathbf{p}+\mathbf{q}/2}+\epsilon _{\mathbf{p}-%
\mathbf{q}/2}}.
\end{equation}%
In the above formula $\nu $ is the volume of the Brillouin zone (BZ), $f^{0}(%
\mathbf{q})=1/[1+\exp ((\epsilon _{q}-\mu )/T)]$ is the free momentum
distribution function and $\epsilon _{q}=-2t\sum_{\alpha }\cos (q_{\alpha }a)
$ is the free particle dispersion relation. The pair correlation functions
are related to the instantaneous structure factors by
\begin{equation}
g_{cc}(\mathbf{r}_{\mathbf{i}})=1+\frac{1}{n}\int_{BZ}\frac{d\mathbf{q}}{\nu
}[S_{cc}(\mathbf{q})-1]\exp (i\mathbf{q}\cdot \mathbf{r}_{\mathbf{i}})
\label{gcc}
\end{equation}%
\begin{equation}
g_{ss}(\mathbf{r}_{\mathbf{i}})=\frac{1}{n}\int_{BZ}\frac{d\mathbf{q}}{\nu }%
[S_{ss}(\mathbf{q})-1]\exp (i\mathbf{q}\cdot \mathbf{r}_{\mathbf{i}})
\label{gss}
\end{equation}%
where $S_{cc,ss}(\mathbf{q})=S_{\sigma \sigma }(\mathbf{q})\pm S_{\sigma
\tilde{\sigma}}(\mathbf{q})$ are the charge and spin component of the
instantaneous structure factor, the spin resolved instantaneous structure
factor being defined by $S_{\sigma \sigma ^{\prime }}(\mathbf{q}%
)=\left\langle n_{\sigma }(\mathbf{q})n_{\sigma ^{\prime }}(\mathbf{q}%
)\right\rangle /n$ with $n_{\sigma }(\mathbf{q})$ the Fourier transform of $%
n_{\mathbf{i}\sigma }$. Finally, the instantaneous structure factors are
connected to the response functions (or susceptibilities) by the
fluctuation-dissipation theorem
\begin{equation}
S_{cc,ss}(\mathbf{q})=-\frac{T}{n}\sum_{\omega _{n}}\chi _{cc,ss}(\mathbf{q}%
,\omega _{n})  \label{fdt}
\end{equation}

\subsection{Particle-hole symmetry for the functional derivatives\label%
{FunctionalDerivatives}}

We are left with the evaluation of the functional derivatives. We first use
the particle-hole (Lieb-Mattis) canonical transformation $c_{\sigma
}(i)=\exp (i\mathbf{Q}\cdot \mathbf{r}_{i})c^{\prime }{}_{{\sigma }%
}^{\dagger }(i)$ (where $\mathbf{Q}=(\pi ,\pi )$) in the definition of the
pair correlation function. In the second step we will use electron-hole
symmetry to evaluate the functional derivative of the pair correlation
function at half-filling. First step: When point $1\ $and point $2$ are
different or when $\sigma \neq \sigma ^{\prime }$ the particle-hole
transformation gives%
\begin{widetext}
\begin{equation}
g_{\sigma \sigma ^{\prime }}(1,2)=\frac{\left\langle n_{\sigma }(1)n_{\sigma
^{\prime }}(2)\right\rangle }{n_{\sigma }(1)n_{\sigma ^{\prime }}(2)}=\frac{%
\left\langle [1-n_{\sigma }^{\prime }(1)][1-n_{\sigma ^{\prime }}^{\prime
}(2)]\right\rangle }{[1-n_{\sigma }^{\prime }(1)][1-n_{\sigma ^{\prime
}}^{\prime }(2)]}=\frac{1-n_{\sigma }^{\prime }(1)-n_{\sigma ^{\prime
}}^{\prime }(2)+n_{\sigma }^{\prime }(1)n_{\sigma ^{\prime }}^{\prime
}(2)g_{\sigma \sigma ^{\prime }}^{\prime }(1,2)}{1-n_{\sigma }^{\prime
}(1)-n_{\sigma ^{\prime }}^{\prime }(2)+n_{\sigma }^{\prime }(1)n_{\sigma
^{\prime }}^{\prime }(2)}  \label{ggp}
\end{equation}%
\end{widetext}%
where $g^{\prime }$ is the pair correlation function of holes at the density 
$n^{\prime }$. Note that $n$ inside the angle brackets are the operator
densities. We abbreviate average densities such as $\left\langle n_{\sigma
}(1)\right\rangle $ by $n_{\sigma }(1)$.

The above equation may look innocuous, but it leads to a very strong
constraint on the pair correlation function and it can be used to find the
exact functional derivative of the pair correlation function at half-filling
where particle-hole symmetry is exact. To obtain this we first take $%
n(1)=1+\epsilon $ and $m(1)=0$ which means $n^{\prime }(1)=1-\epsilon $ and $%
m^{\prime }(1)=0$. Note that $n\left( 2\right) $ is unchanged. Using the
fact that $n_{\uparrow ,\downarrow }(i)=[n(i)\pm m(i)]/2$ one obtains 
\begin{equation}
g_{\uparrow \sigma }(1,2)=\frac{\frac{\epsilon }{2}+\frac{1-\epsilon }{4}%
g_{\uparrow \sigma }^{\prime }(1,2)}{\frac{1+\epsilon }{4}}.
\end{equation}%
When there is exact electron-hole symmetry, such as at half-filling, the
electron pair correlation function at density $n$ is equal to the hole pair
correlation function at the same density, so we deduce from the above
equation that 
\begin{equation}
\frac{\delta g_{\uparrow \sigma }(1,2)}{\delta n(1)}=[1-g_{\uparrow \sigma
}(1,2)].
\end{equation}%
The summation of last equation for two different values of $\sigma $ gives
us one of the functional derivatives 
\begin{equation}
\frac{\delta g_{s\sigma }(1,2)}{\delta n(1)}=[1-g_{cc}(1,2)].
\end{equation}%
Using the same trick we can also show that 
\begin{equation}
\frac{\delta g_{\uparrow \downarrow }(1,1)}{\delta n(1)}=2[1-g_{\uparrow
\downarrow }(1,1)].  \label{dg_up_dwn_/dn}
\end{equation}%
To evaluate the functional derivative respect to magnetization, it suffices
to assume that $n(1)=1$ and $m(1)=\epsilon $ . It is then easy to show that 
\begin{equation}
\frac{\delta g_{s\sigma }(1,2)}{\delta m(1)}=[1-g_{cc}(1,2)].
\end{equation}%
Using the same trick one can also show, as mentioned before, that at half
filling 
\begin{equation}
\frac{\delta g_{\sigma \sigma ^{\prime }}(1,2)}{\delta (n,m)(3)}=0
\label{Vanishing_fd}
\end{equation}%
when site $3$ is different from both $1\;$and$\;2$. Although there is no
strict particle-hole symmetry away from half-filling, it is reasonable to
assume that at weak to intermediate coupling the Pauli principle forces only
electrons close to the Fermi surface to be relevant. In that case, the
dispersion relation can be linearized and particle-hole symmetry can be
safely assumed for the pair correlation functions $g_{\sigma \sigma ^{\prime
}}(i,j).$ The comparisons with QMC calculations presented later suggest that
it is indeed a good approximation.

\subsection{Self-consistency and discussion of Extended Two-Particle
Self-Consistent Approach (ETPSC)\label{Def_ETPSC}}

Substituting the above equations for the functional derivatives in the
expression for the irreducible vertices $U_{cc}\left( \mathbf{q}\right) ,$
Eq.(\ref{Ucc}), and $U_{ss}\left( \mathbf{q}\right) ,$ Eq.(\ref{Uss}), one
can compute the charge and spin susceptibilities Eqs.(\ref{chi_cc},\ref%
{chi_ss}) given the three unknown pair correlation functions $g_{\sigma 
\tilde{\sigma}}(1,1),$ $g_{cc}(1,1+a)$ and $g_{ss}(1,1+a).$ We do not need $%
g_{\sigma \sigma }(1,1)$ because we know from the Pauli principle that it
vanishes. The three unknowns can be determined self-consistently as follows.
Use the fluctuation-dissipation theorem Eq.(\ref{fdt}) to obtain $S_{cc,ss}(%
\mathbf{q})$ from the susceptibilities and then substitute in the
definitions Eqs.(\ref{gcc},\ref{gss}) of the pair correlation functions in
terms of $S_{cc,ss}(\mathbf{q}).$ From this procedure, we obtain four
equations for three unknowns since a look at the left-hand side of Eqs.(\ref%
{gcc},\ref{gss}) tells us that we can compute $g_{cc}(0)=g_{cc}(1,1),$ $%
g_{ss}(0)=g_{ss}(1,1)$, as well as $g_{cc}(\mathbf{a})=g_{cc}(1,1+a)$ and $%
g_{ss}(\mathbf{a})=g_{ss}(1,1+a).$ The problem is thus over determined. To
be consistent with the initial assumption $g_{\sigma \sigma }(1,1)=0$ from
the Pauli principle, the self-consistent solution of these four equations
have to lead to $g_{\sigma \sigma }\left( 0\right) =g_{cc}\left( 0\right)
+g_{ss}\left( 0\right) =0.$ In ordinary TPSC for the on-site Hubbard model,
one does not have a separate equation for $\delta g_{\sigma \tilde{\sigma}%
}(1,1)/\delta n(1)$ that enters $U_{cc}.$ Instead it is determined by
enforcing that $g_{\sigma \sigma }\left( 0\right) =0$ in the sum rules Eqs.(%
\ref{gcc},\ref{gss}). In the present extension of TPSC to near-neighbor
interaction we have an expression for that functional derivative, Eq.(\ref%
{dg_up_dwn_/dn}) so in general the four equations obtained from the
sum-rules Eqs.(\ref{gcc},\ref{gss}) do not satisfy the on-site Pauli
principle $g_{cc}\left( 0\right) +g_{ss}\left( 0\right) =g_{\sigma \sigma
}\left( 0\right) =0,$ which is assumed in their derivation.

At this point there are three possible ways to proceed. (a) Keep the
equation (\ref{gss}) for $g_{ss}\left( 0\right) $ and drop that for $%
g_{cc}\left( 0\right) $ in Eq.(\ref{gcc}) (b) Do the reverse procedure, i.e.
keep the equation (\ref{gcc}) for $g_{cc}\left( 0\right) $ and drop that for 
$g_{ss}\left( 0\right) $ in Eq.(\ref{gss})$.$ (c) In the original spirit of
TPSC, do not use our estimate of the functional derivative $\delta g_{\sigma 
\tilde{\sigma}}(1,1)/\delta n(1)$ Eq.(\ref{dg_up_dwn_/dn}) but determine the
value of 
\begin{equation*}
g_{\sigma \tilde{\sigma}}(1,1)+n\frac{\delta g_{\sigma \tilde{\sigma}}(1,1)}{%
\delta n(1)}
\end{equation*}%
entering the charge vertex Eq.(\ref{Ucc}) by solving the four equations for $%
g_{cc}(0)=g_{cc}(1,1),$ $g_{ss}(0)=g_{ss}(1,1)$, $g_{cc}(\mathbf{a}%
)=g_{cc}(1,1+a)$ and $g_{ss}(\mathbf{a})=g_{ss}(1,1+a)$ that follow from
Eqs.(\ref{gcc},\ref{gss}). The latter approach, (c), reduces to TPSC when
the nearest-neighbor interaction $V$ vanishes and gives the best
approximation in that case. With this procedure, $g_{\sigma \sigma }\left(
0\right) =0$ is still true. However, comparisons with Quantum Monte Carlo
calculations also suggest that in the presence of $V,$ procedure (a) is a
better approximation. When $V$ vanishes, one recovers TPSC for the spin
fluctuations. These fluctuations are anyway dominant, while the charge
fluctuations, different from TPSC, are reduced compared with the
non-interacting case. With the latter procedure, (a), the sum rule $%
g_{cc}\left( 0\right) +g_{ss}\left( 0\right) =0$ that follows from the Pauli
principle $\left\langle n_{\sigma }^{2}\right\rangle =\left\langle n_{\sigma
}\right\rangle $ is not satisfied, but the equation that allows us to find $%
U_{ss}$ when $V=0$ does take into account the Pauli principle and gives
exactly the same result as TPSC.

Following the same reasoning as in Appendix A of Ref.(%
\onlinecite{Vilk2}%
), we also find that our approach satisfies the $f$ sum rule and the
Mermin-Wagner theorem. Following the lines of Appendix B of Ref.(%
\onlinecite{Bahman}%
), one can also develop a formula for the self-energy at the second step of
the approximation that satisfies the relation between $\Sigma G$ and
potential energy. The latter is a kind of consistency relation between one-
and two-particle quantities.


\section{Numerical results}

\label{sec4}

\subsection{Accuracy of the approach}

Following the procedure (a) explained at the end of the previous section, we
present in the first six figures various results that allow us (i) to
discuss some of the physics contained in the pair correlation functions and
(ii) to establish the validity of the approach and the sources of error by
comparison with benchmark Quantum Monte Carlo calculations and with other
approximations. All the calculations are for the two-dimensional square
lattice. In most cases, unless otherwise specified, we will consider how the
physics changes when $V$ is varied for $U=4$ and $n=1$ fixed. We are using
units in which $\hbar ,k_{B},t$ and lattice spacing $a$ are all taken to be
unity. When the spin or charge components of the response functions are
growing rapidly at low temperature towards very large values, we stop
plotting the results. The last two figures of the paper contain general
crossover diagrams for the various forms of charge and spin orders,
neglecting pairing channels.

In Figs.~1 and 2, we first discuss the case $V=0.$ This is necessary since
our approximation (for charge fluctuations) is different from TPSC even in
that case. Fig. 1 compares results of our numerical calculations with QMC,
TPSC and FLEX.
\begin{figure}[tbp]
\begin{center}
\includegraphics[scale=0.5]{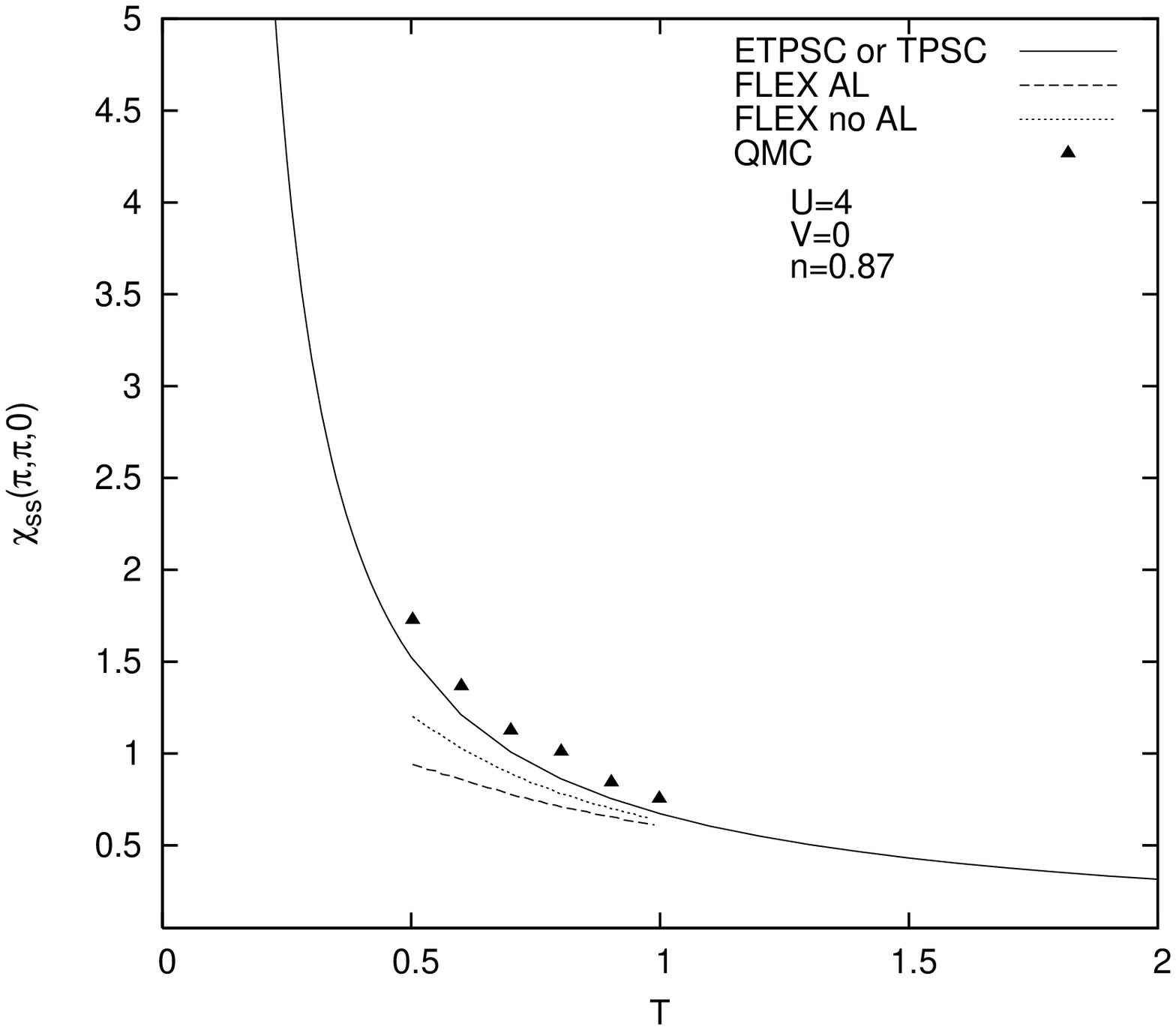} %
\includegraphics[scale=0.5]{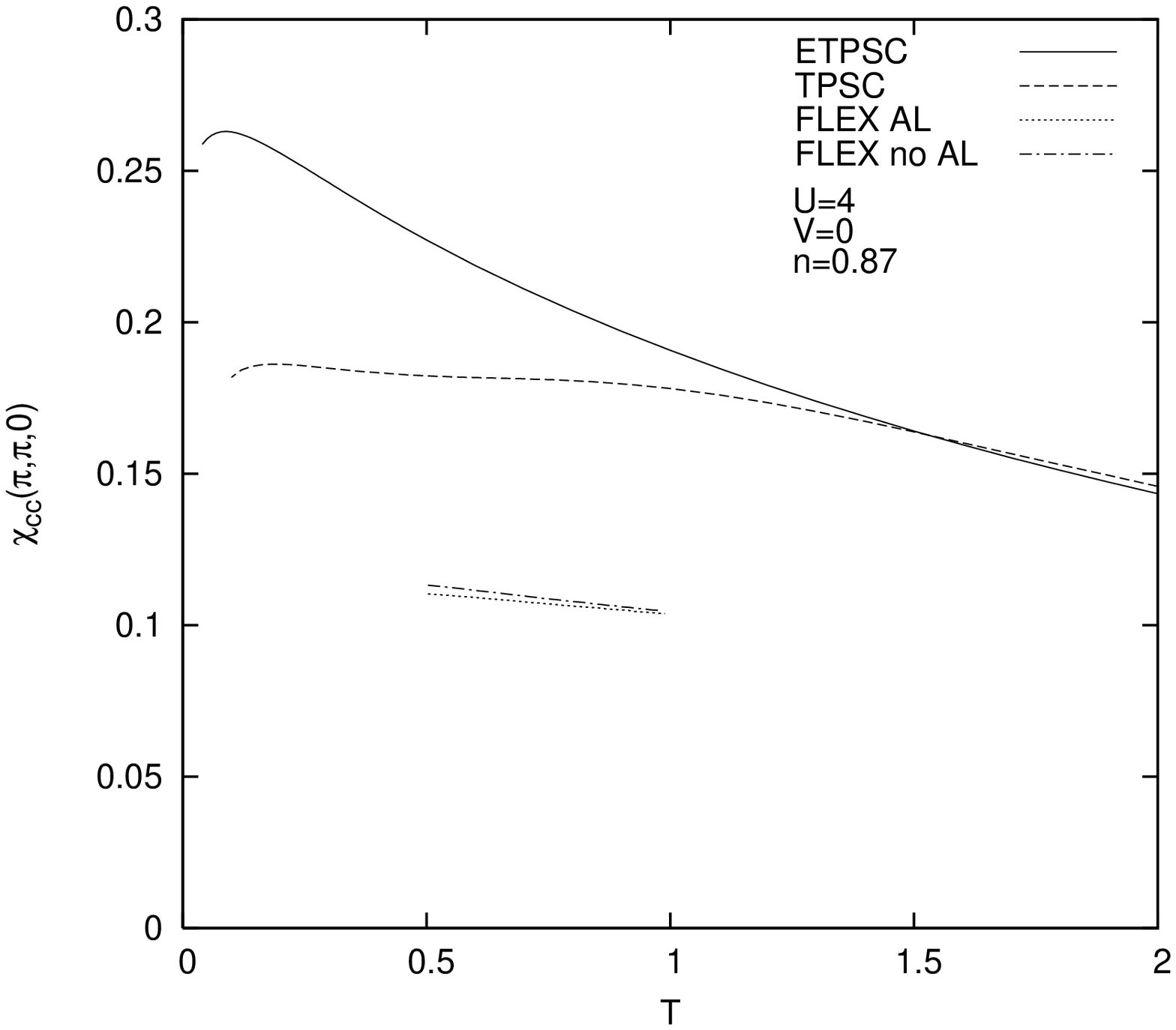}
\end{center}
\caption{The spin (left panel) and charge (right panel) components of static
(zero frequency) susceptibilities at $n=0.87$, $U=4$ and $V=0$ as a function
of the temperature. FLEX and QMC results are from Ref.
\onlinecite{BickersWhite:1991}%
.}
\end{figure}
We depict the charge (right panel) and spin (left panel) components of the
static (zero frequency) susceptibilities at $n=0.87$, $U=4$ and $V=0$ as a
function of temperature. Consider the spin susceptibility on the left. As we
can see, the FLEX approximation underestimates the spin susceptibility. The
response function within TPSC stays slightly below benchmark QMC results
although the TPSC result for structure functions \cite{Vilk1} (not presented
here) are almost on top of QMC results. The results from the ETPSC method
developed in this paper coincide with the TPSC results for the spin
component. The charge susceptibility differs between TPSC and ETPSC as
illustrated on the right panel. We do not have QMC results for the charge
susceptibility but the charge structure functions for TPSC and QMC are quite
close, \cite{Vilk1} although not as much as in the case of the spin
structure functions. We can thus surmise that FLEX underestimates the charge
susceptibility. In ETPSC, the charge susceptibility is larger than in TPSC.%
\begin{widetext}%

\begin{figure}[tbp]
\begin{center}
\includegraphics[scale=0.4]{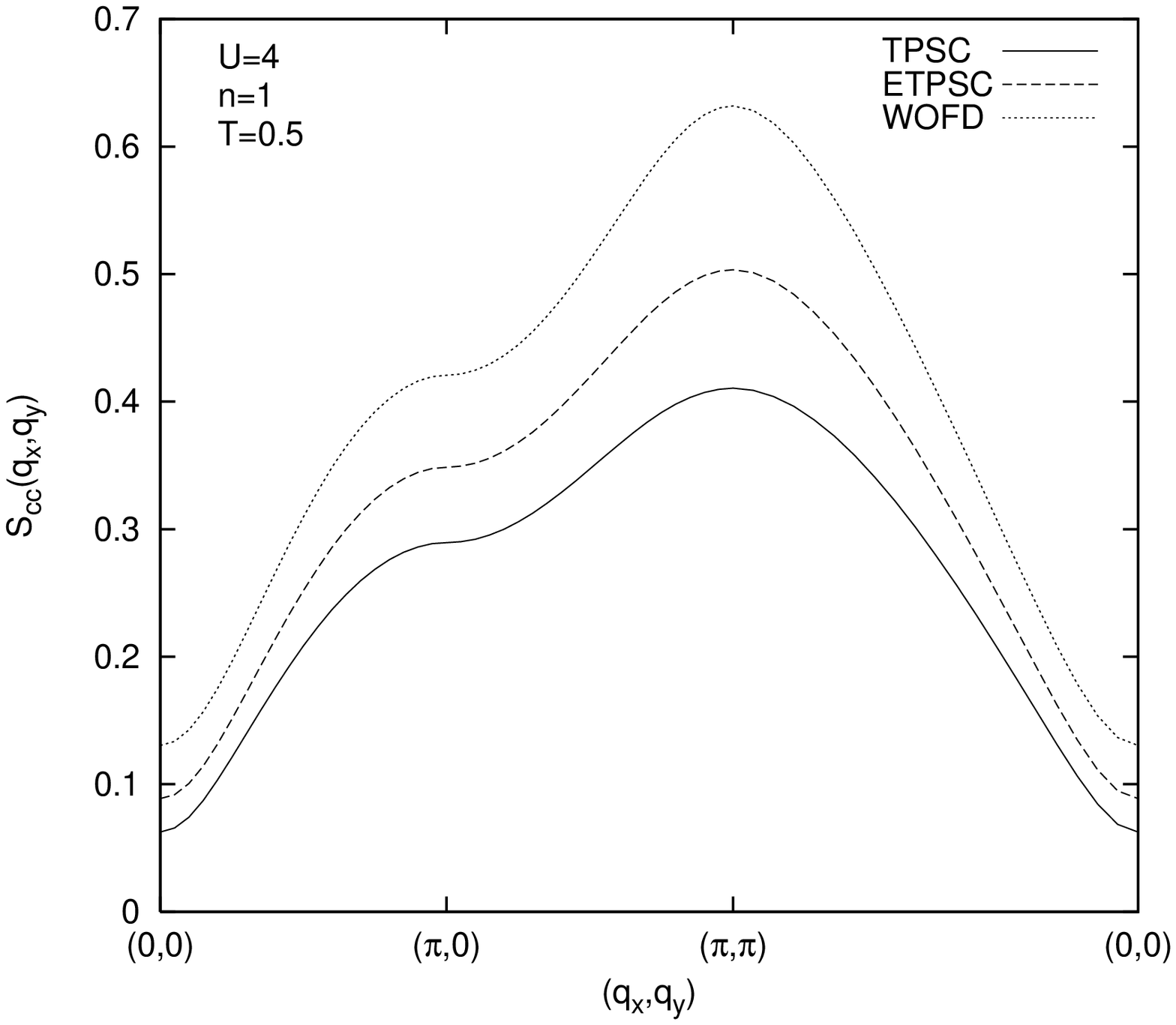} %
\includegraphics[scale=0.4]{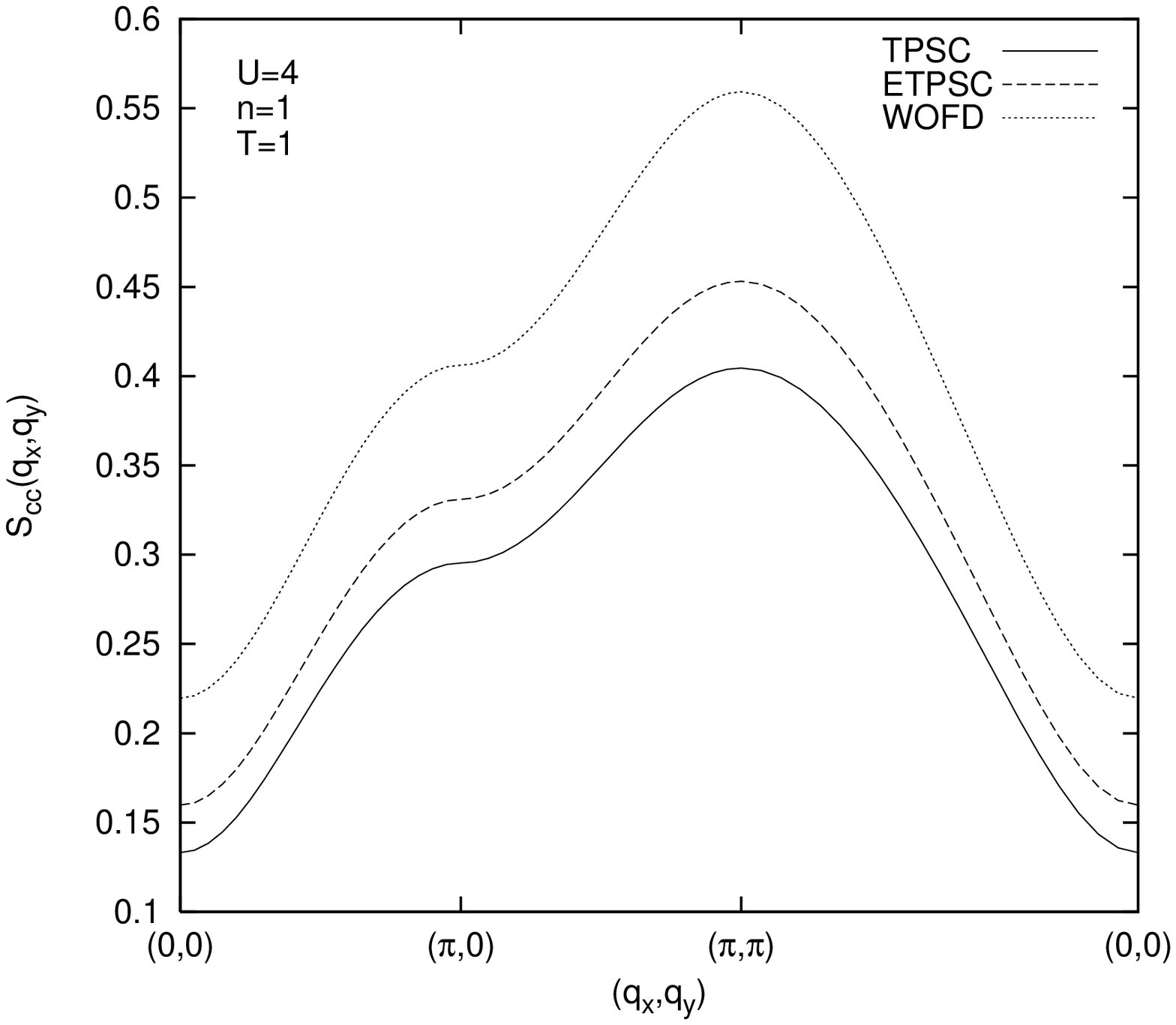} %
\includegraphics[scale=0.4]{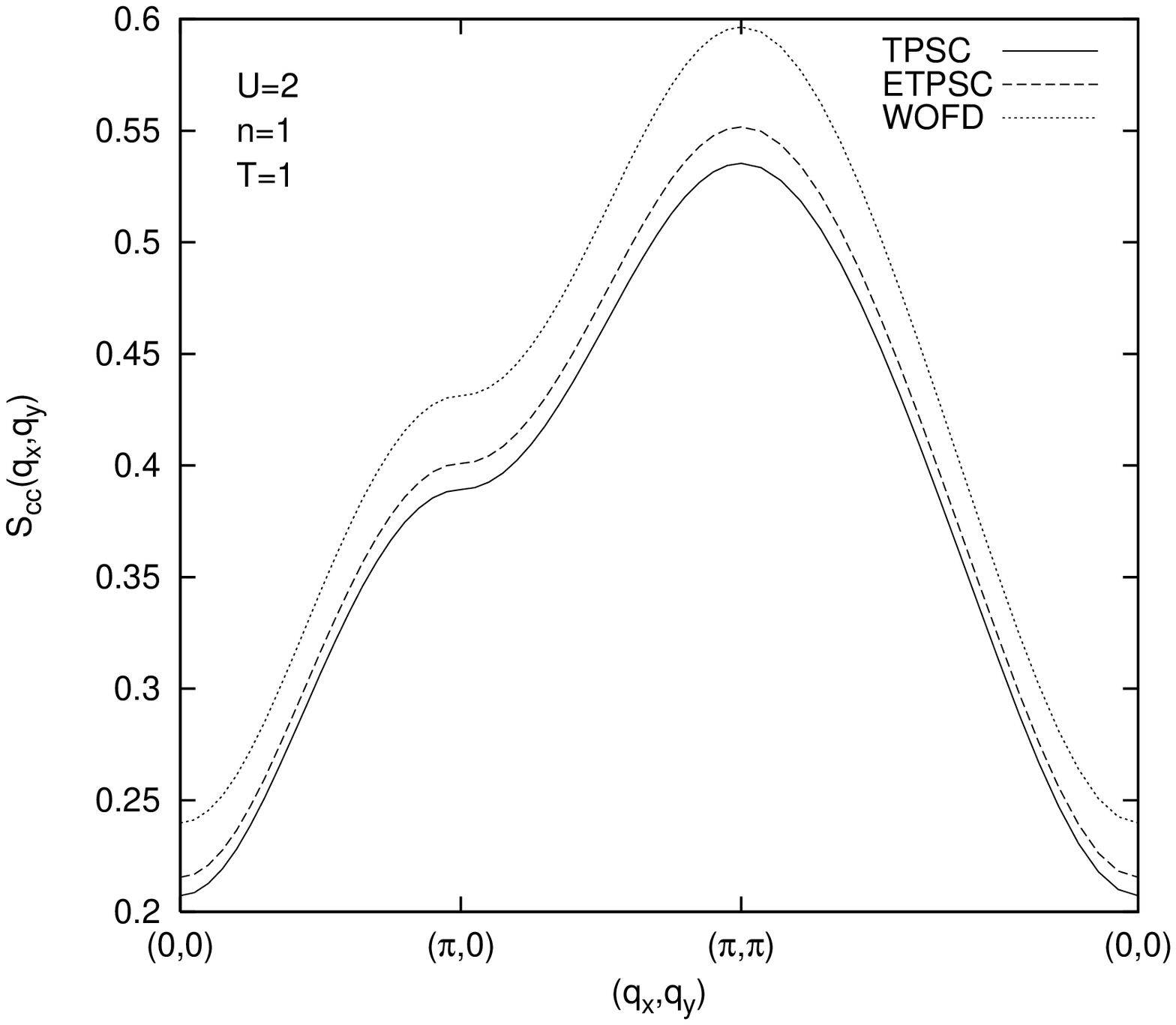} %
\includegraphics[scale=0.4]{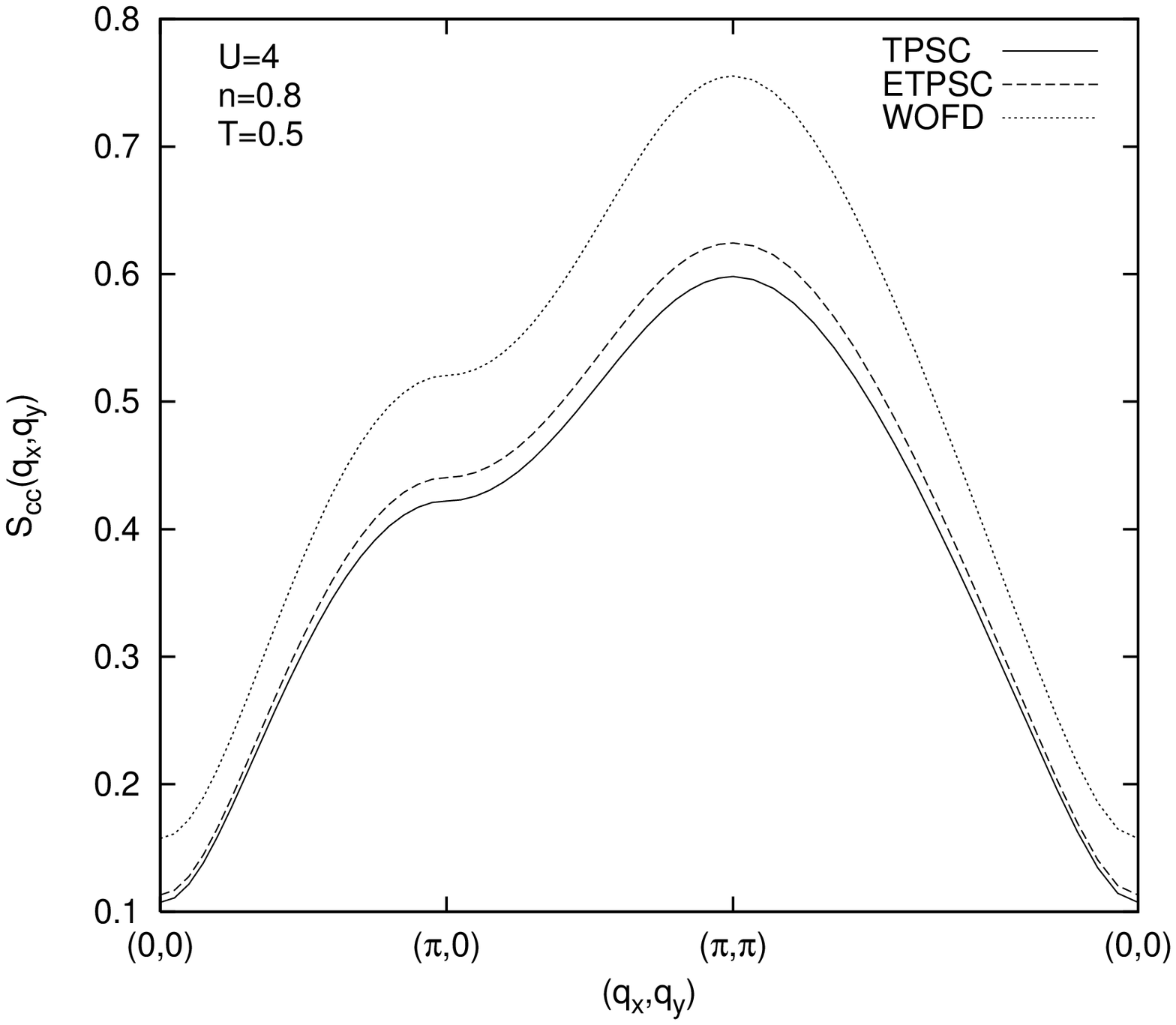}
\end{center}
\caption{Comparison of the ETPSC results for the instantaneous charge
structure factor with TPSC and with WOFD.}
\end{figure}

\end{widetext}%
Even though the charge fluctuations are less important than the spin fluctuations when $V
$ vanishes, it is instructive to consider them to gauge the effect of our
two approximations, namely factorization with a local field factor Eq.(\ref%
{Sigma+}), approximate and estimate of the functional derivatives in Sec. \ref {FunctionalDerivatives}. We compare in Fig. 2 the charge component of the
TPSC instantaneous structure factor with ETPSC, where the functional
derivative terms are evaluated, and with our previous approach \cite{Bahman}
(WOFD stands for Without Functional Derivatives) that neglects completely
these functional derivatives. The
temperatures are relatively high but they are low enough to show
discrepancies between the various approaches (TPSC essentially coincides
with QMC results in this range of physical parameters). The importance of
the functional derivatives is obvious from all figures in the different
panels since ETPSC always agrees better with TPSC than WOFD. Given the
assumption that the pair correlation functions are functionals of the
density (Eq.(\ref{Functional_of_density})), our evaluation of the functional
derivatives at half-filling is exact. Hence the three panels for $n=1$ in
Fig.2 are particularly instructive. They suggest that the main difference
between ETPSC and TPSC is due either to the factorization Eq.(\ref{Sigma+}),
or to the assumption Eq.(\ref{Functional_of_density}) that the pair
correlation functions are functionals of the density only.  In other words,
the difference between ETPSC and TPSC suggests the limits of the
factorization assumption. TPSC seems to compensate the limitations of the
factorization by fixing the functional derivative term with the Pauli
principle. One can also find the degree of violation from the Pauli sum-rule
by comparison of our data with TPSC. This is the main source of error in our
calculation and it is more pronounced at larger $U,$ lower $T$ and near
half-filling. The bottom-right panel shows that ETPSC is in better agreement
with TPSC away from half filling, an encouraging result.

\begin{figure}[tbp]
\begin{center}
\includegraphics[scale=0.5]{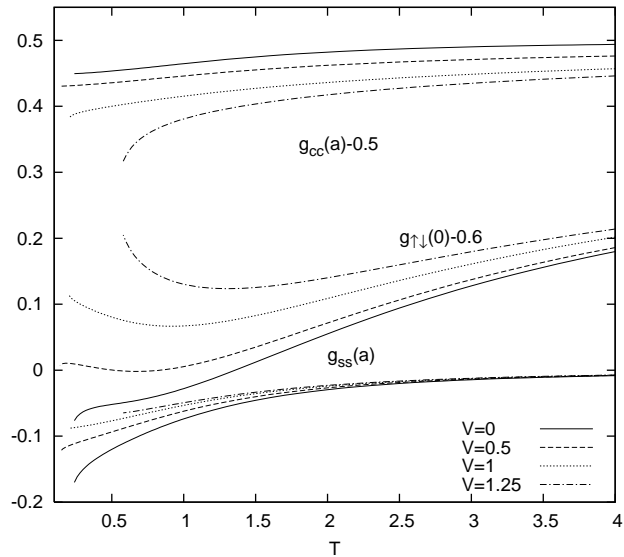}
\end{center}
\caption{The pair correlation functions as a function of temperature at $n=1$%
, $U=4$, for various values of the nearest-neighbor interaction $%
V=0,\;0.5,\;1\;\mathrm{and}\;1.25.$}
\end{figure}

To illustrate the physics associated with the nearest-neighbor interaction $%
V,$ we show in Fig. 3 the variation of $g_{\sigma \tilde{\sigma}}(0)$ and $%
g_{cc,ss}(a)$ with temperature for various values $V=0,\;0.5,\;1\;$and$%
\;1.25 $. We first notice that $g_{\sigma \tilde{\sigma}}(0)$ at $V=0$ has a
sharp decrease around $T\approx 0.3,$
which means that the probability for finding two particles at the same
place sharply decreases with decreasing the temperature around that point. $%
g_{ss}(a)$ also becomes more negative around the same temperature, which
means that the probability of finding of two electrons at a distance $a$ and
with opposite spins is higher than finding them with the same spins. These
two results are a consequence of the tendency towards Spin-Density Wave
(SDW) order at zero temperature. The decrease of $g_{\sigma \tilde{\sigma}%
}(0)$ around $T\approx 0.3$ signals then entry into the renormalized
classical regime, as discussed in Ref.
\onlinecite{Vilk2}%
. In the charge sector, $g_{cc}(a)$ does not show any strong change in the
low temperature limit, which means there are no strong Charge Density Wave
(CDW) fluctuations at $V=0$. We observe that this feature changes as we
increase $V$, which shows that SDW fluctuations are depressed in favor of
CDW fluctuations. This can easily be observed at $V=1.25,$ which shows that $%
g_{cc}(a)$ has a sharp decrease around $T\approx 0.5$ and instead $g_{\sigma
\tilde{\sigma}}(0)$ has a sudden increase. On the other hand $g_{ss}(a)$
dose not show any decrease around that temperature. These results together
show for $V=1.25$ the buildup of CDW rather than SDW fluctuations as
temperature decreases. We stress that all these low temperature behaviors
are non-perturbative. The high temperature behavior where the quantities
vary smoothly can be understood perturbatively.

\begin{figure}[tbp]
\begin{center}
\includegraphics[scale=0.5]{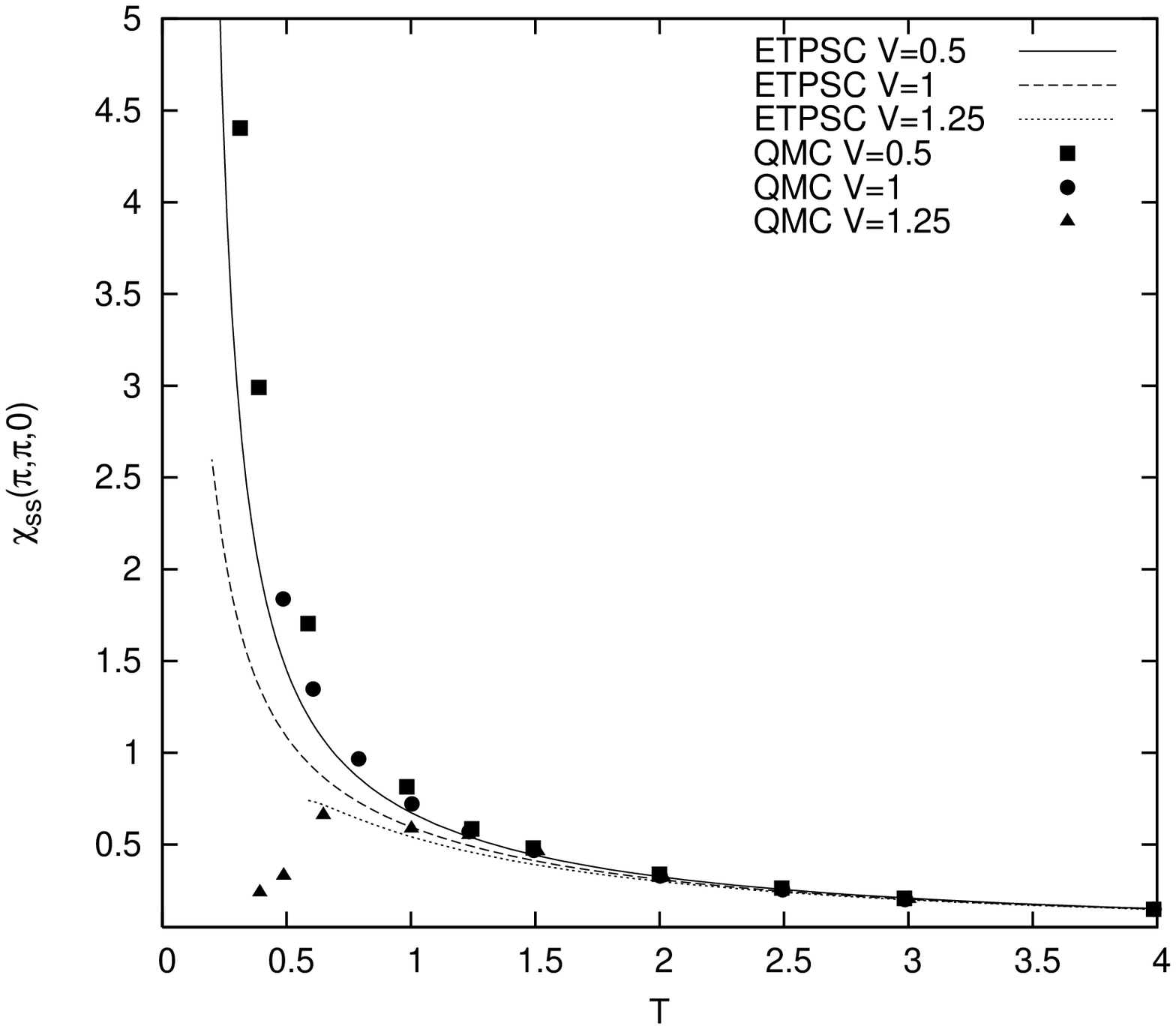} %
\includegraphics[scale=0.5]{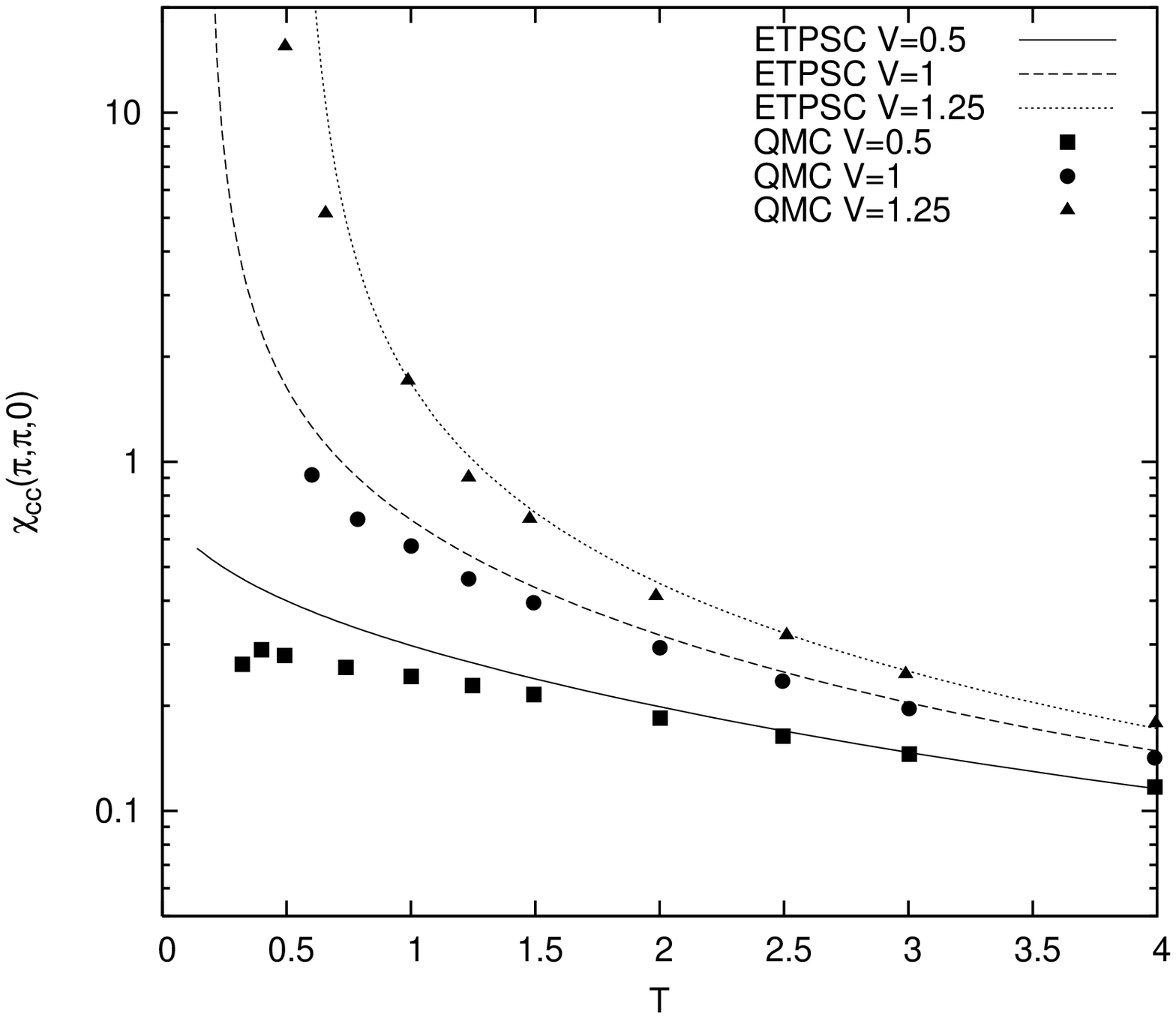}
\end{center}
\caption{The spin (left panel) and charge (right panel) zero-frequency
response functions at $n=1$, $U=4$, $V=0.5,\;1\;\mathrm{and}\;1.25$. Points
are the QMC results \protect\cite{Zhang} and lines are the ETPSC results.}
\end{figure}

To gauge the accuracy of our approach at finite $V$, we now compare ETPSC
results with benchmark QMC calculations \cite{Zhang} for various values of $%
V=0.5,\;1\;$and$\;1.25$. We plot in Fig.~4 the static (zero frequency) spin
and charge response functions evaluated at $Q=(\pi ,\pi )$ as a function of
temperature along with the QMC results in the left and right panels
respectively. As is clear from the figure, there is a good agreement between
our results and the QMC results. As $V$ increases, the charge response is
largest and for that quantity the agreement improves at larger $V$. A simple
comparison of this figure with the results of our previous paper \cite%
{Bahman} shows the importance of the functional derivative terms and the
accuracy of their evaluation. One also observes that the spin component of
the response function exhibits a rapid decrease at low temperature limit
when $V=1.25.$ An analogous behavior occurs at $V=0$ (not presented here)
for the charge response function. This feature is a result of the Pauli
principle, which implies that in the paramagnetic state charge and spin
fluctuations cannot both increase \cite{Vilk2}. The small deviation of the
charge response (or instantaneous structure factor) at $V=0.5$ can be
removed by implementing this sum-rule \cite{Bahman}. We also observe that
the spin component of the response functions at different values of $V$ all
tend to the same value at sufficiently high temperature, which simply means
that $g_{ss}(1,1+a)$ and $\frac{\delta g_{s}(1,1+a)}{\delta m(1)}$ are
negligible in this limit. This is the only region where ETPSC coincides with
Random Phase Approximation (RPA) results and the only place where RPA is
valid. The effects of the above-mentioned terms become very important at low
temperature and that is one of the sources of failure of RPA.
\begin{figure}[tbp]
\begin{center}
\includegraphics[scale=0.5]{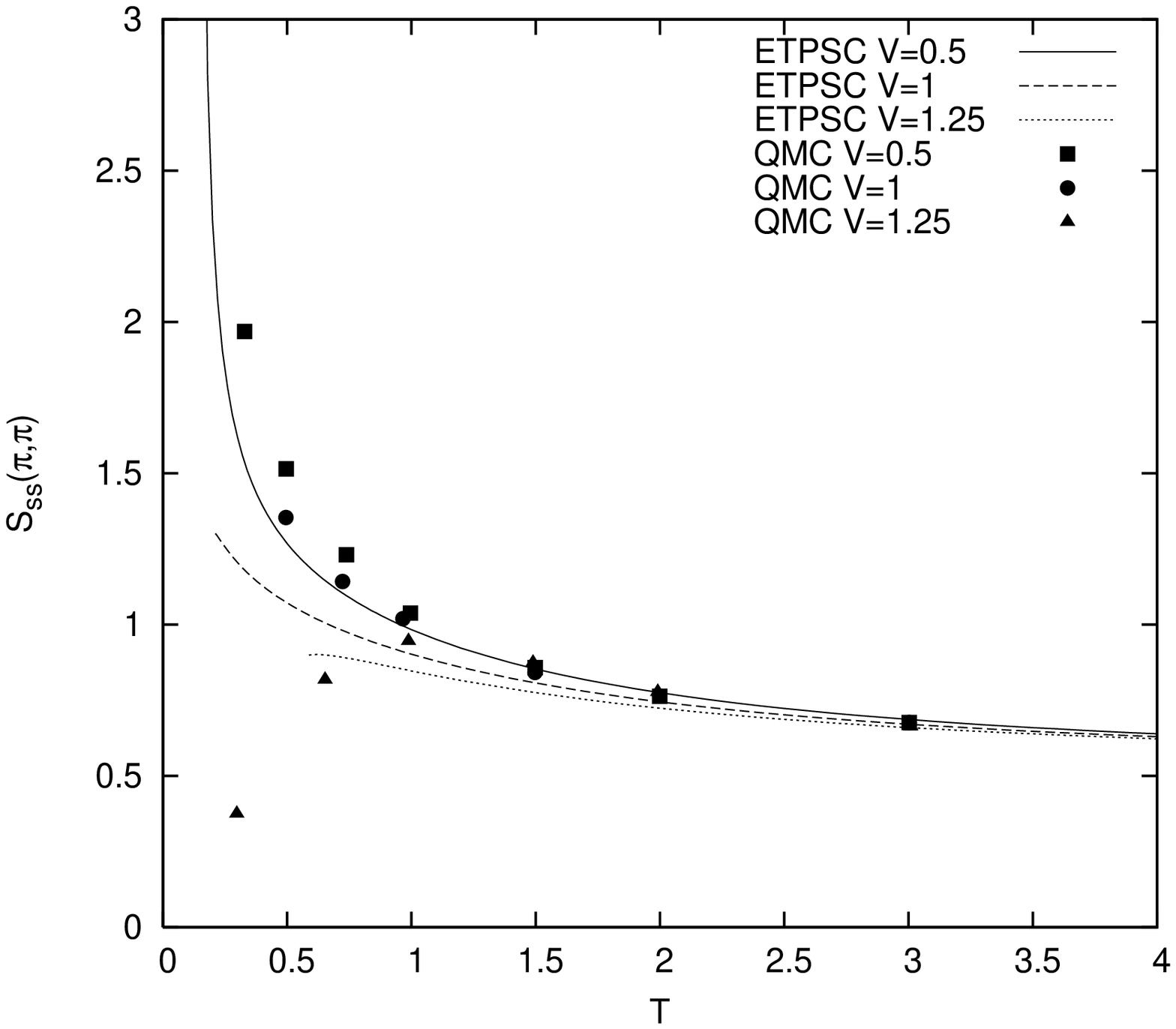} %
\includegraphics[scale=0.5]{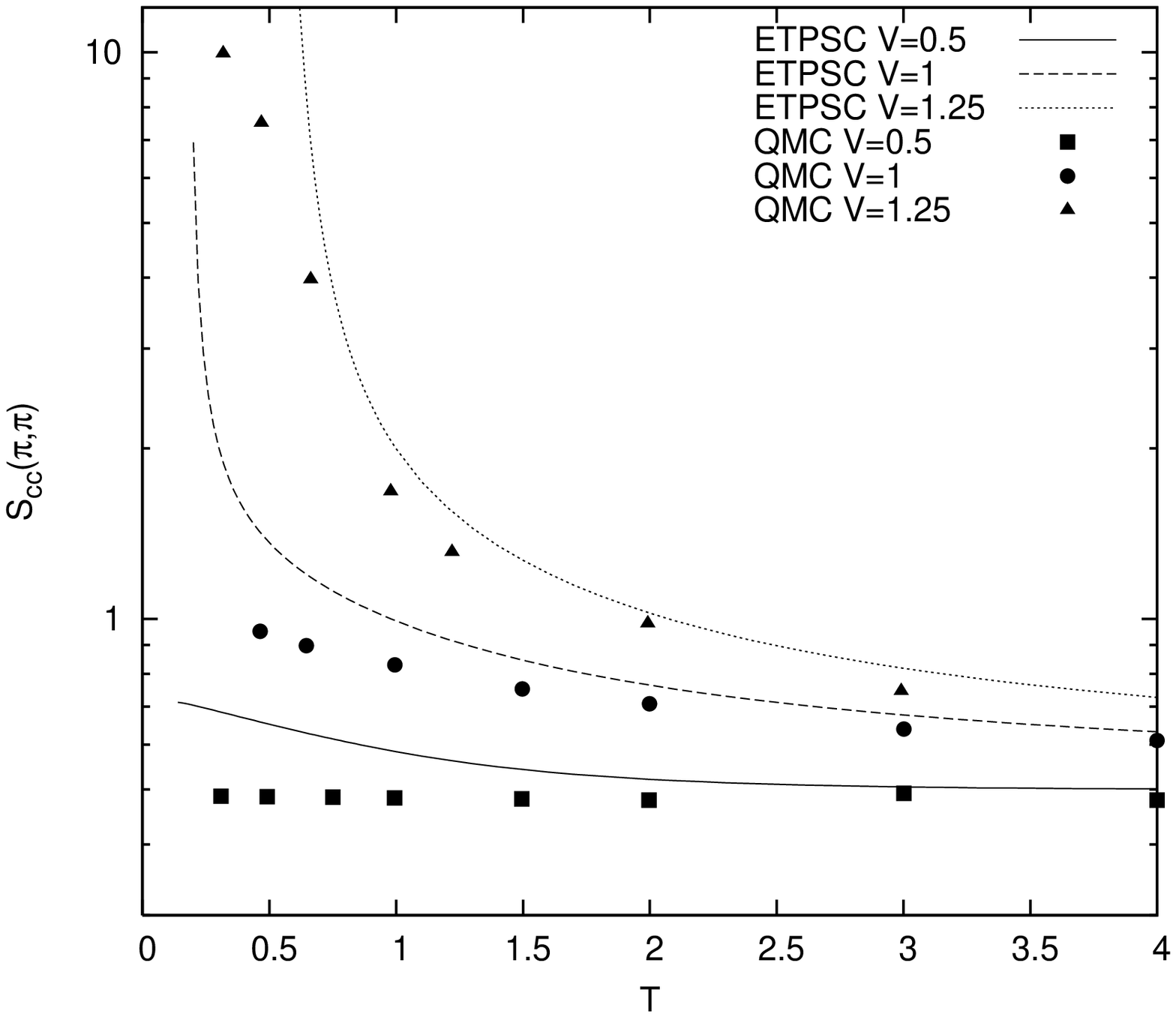}
\end{center}
\caption{The spin (left panel) and charge (right panel) static structure
factors at $n=1$, $U=4$, $V=0.5,\;1\;$and$\;1.25$. Symbols are the QMC
results \protect\cite{Zhang} and curves are the ETPSC results.}
\end{figure}

Fig.~5 shows the instantaneous spin and charge structure factors at $Q=(\pi
,\pi )$ and $V=0.5,\;1\;$and$\;1.25$ as a function of temperature. All the
features are similar to what was mentioned in Fig.~4 except that the
agreement is less satisfactory. This suggests that our approach with
constant vertices is best for low frequency response functions. The
structure factors contain contributions from all frequencies and we know for
example that the irreducible vertices should become equal to the bare
interaction at large frequency \cite{Vilk2}. The ETPSC approach that we are
using, described in procedure (a) Sec. \ref{Def_ETPSC}, gives better
agreement with QMC at higher values of $V$. In the case of the charge
structure factor, procedure (c) compares more favorably with QMC for small $%
V $.

\begin{figure}[tbp]
\begin{center}
\includegraphics[scale=0.5]{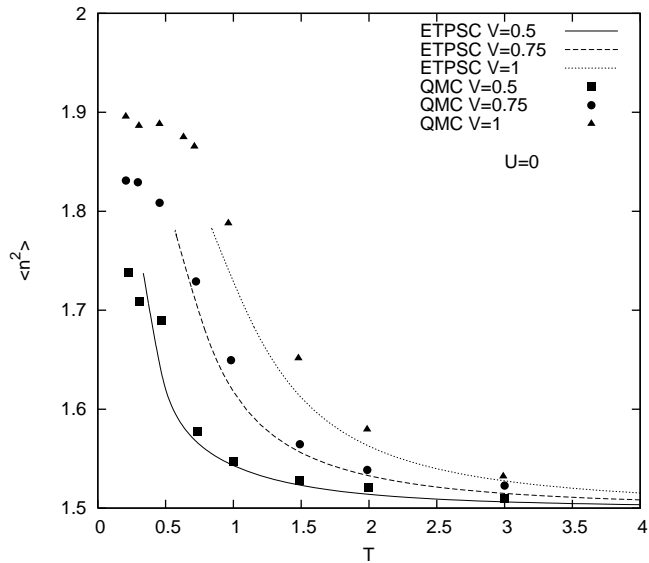}
\end{center}
\caption{The density-density correlation function in term of the temperature
at $V=0.5,\;0.75$ and $1$ for $U=0$. Symbols are the QMC results
\protect\cite{Zhang} and curves are the ETPSC results.}
\end{figure}

Fig.~6 shows the density-density correlation function ($\left <n^2\right >=\left <(n_{\downarrow}+n_{\uparrow})^2\right >$) as a function of temperature
for $V=0.5,\;0.75$ and $1$ in the extreme case $U=0$. The agreement with QMC
is quite remarkable. In this case, we used procedure (b) in Sec. \ref%
{Def_ETPSC} that fixes $g_{\sigma \tilde{\sigma}}(0)$ from the charge rather
than the spin component of the response functions. The difference with
procedure (a) is small, but in this limiting case where charge fluctuations
dominate, procedure (b) is the one that agrees best with QMC.

\subsection{Crossover temperatures towards renormalized classical regimes}

The Mermin-Wagner theorem prevents breaking of a continuous symmetry in two
dimensions. Instead, as temperature decreases one observes a crossover
towards a regime where the characteristic fluctuation frequency of a
response function at some wave vector becomes less than temperature (in
dimensionless units). This is the so-called renormalized-classical regime.
In that regime, the correlation length grows exponentially until a
long-range ordered phase is reached at $T=0.$ The wave vector of the
renormalized classical fluctuations does not depend only on the
non-interacting susceptibility and it does need to be guessed, it is
self-determined within ETPSC. Based on previous experience with TPSC and on
numerical results of the previous section, ETPSC will fail below the
crossover temperature when the correlation length becomes too large, but the
crossover temperature itself is reliable.

We present in Figs.~ 7 and 8 the crossover temperature as a function of $U$
and $V$ for $n=1$ and $n=0.75$ respectively. Even though we display results
for negative $U$ and $V$ as well, other instabilities, in particular in
pairing channels, can dominate when one of the interactions is negative. One
would need to generalize ETPSC for negative interactions along the lines of
Refs.
\onlinecite{AllenTremblay,AllenKyung}%
, but this has not been done yet. The results for negative interactions are
given for illustrative purposes only.

The crossover temperatures in Figs. 7 and 8 are obtained from the condition $%
\chi _{cc}(q_{x},q_{y},0)/\chi _{0}(q_{x},q_{y},0)=10$ or $\chi
_{ss}(q_{x},q_{y},0)/\chi _{0}(q_{x},q_{y},0)=10,$ where $q_{x}$ and $q_{y}$
refer to the wave vector where the response function has a maximum. These
quantities are proportional to $\xi ^{2},$ the square of the corresponding
correlation length. The characteristic spin fluctuation frequency scales
like $\xi ^{-2}.$ When the crossover temperature is reached, $\xi $
increases so rapidly that a choice different from $10$ does not change very
much the estimate and can increase sensibly the computation time. A more
detailed discussion on the crossover temperature can be found in Ref. 
\onlinecite{Bahman}%
.%
\begin{widetext}%

\begin{figure}[tbp]
\begin{center}
\includegraphics[scale=0.8]{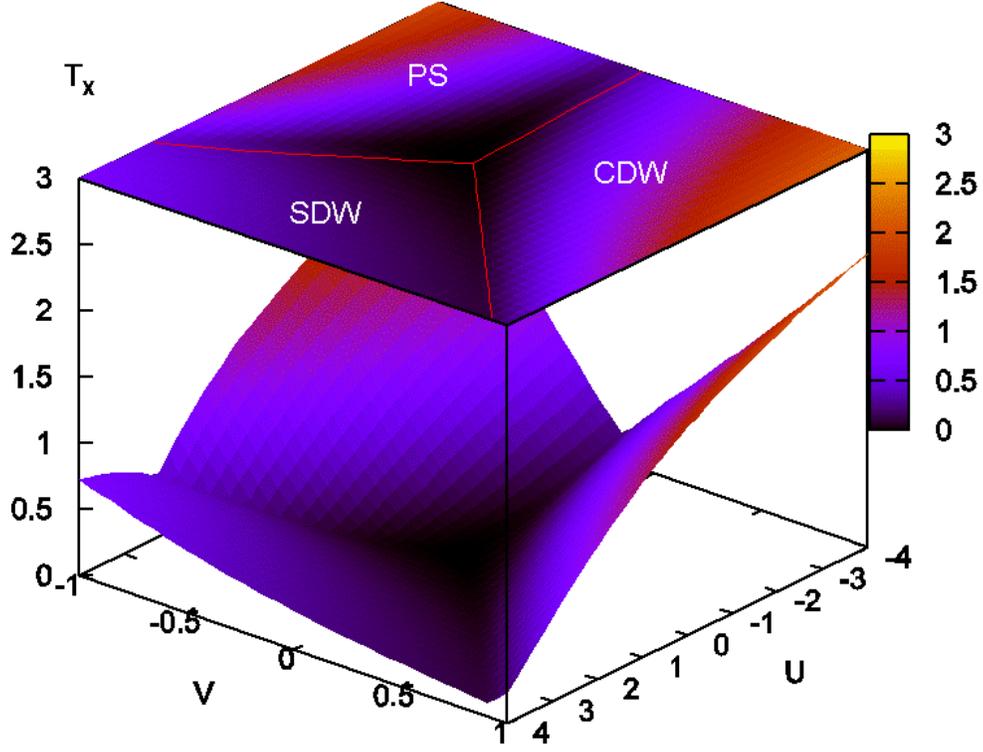}
\end{center}
\caption{The crossover temperature at $n=1$. The red lines separate $\left( 
\protect\pi ,\protect\pi \right) $ SDW (left), $\left( \protect\pi ,\protect%
\pi \right) $ CDW (right) and PS (on top). Results for negative values of
the interaction strength are given for illustrative purposes only. Pairing
instabilities may dominate in these cases.}
\end{figure}

\end{widetext}%
The red lines in Fig.~7 separate three different fluctuation regimes, namely 
$\left( \pi ,\pi \right) $ SDW (on the left), $\left( \pi ,\pi \right) $ CDW
(on the right) and phase separation (PS) (this is defined by the growth of
the zero frequency charge response function at $q_{x}=0$ and $q_{y}=0$). It
is noteworthy that the boundary between the SDW and CDW crossover is closely
approximated by $U\approx 4V$ (for positive $U$ and $V$), which is in good
agreement with previous results \cite%
{Zhang,Onari,Oles,Ohta:1994,Yan:1993,Bari:1971}. Theory that includes
collective modes beyond mean-field also shows deviations from $U=4V$ at weak
to intermediate coupling. \cite{Yan:1993} Since there are four neighbors
which, through $V$, favor charge order while $U$ favors spin order, the
crossover at $U\approx 4V$ is expected. In fact, the result $U=zV,$ where $z$
is the number of nearest-neighbors, seems to hold quite generally in various
dimensions at $T=0$. \cite{Oles,Fourcade,Micnas} Slightly away from
half-filling $\left( n=0.9\right) $ and with a small second-neighbor
hopping, the MF approximation \cite{Murakami} predicts a coexistence region
for SDW, CDW and ferromagnetism in the large $U$ and $V$ region of parameter
space. This is likely to be an artefact due to an inaccurate estimation
of the spin fluctuations in MF when $V\neq 0$. One can see the weakness of
this approach by setting $g_{ss}=0$ and ignoring its functional derivative
in the equation for the spin susceptibility Eq.~(\ref{chi_ss}). These terms
are the main source of the suppression of the SDW so that one obviously
overestimates the spin effects by ignoring these terms when $V$ is large.
One can easily notice that the charge and spin fluctuations appear at
relatively large temperature over most of parameter space. In addition,
negative $U$ greatly increases charge fluctuations. These will compete with
pairing fluctuations that we have not taken into account here. At positive $U
$ and $V,$ $d-$wave superconductivity is not likely to occur in regions
other than that which is nearly black in the figure.%
\begin{widetext}%

\begin{figure}[tbp]
\begin{center}
\includegraphics[scale=0.8]{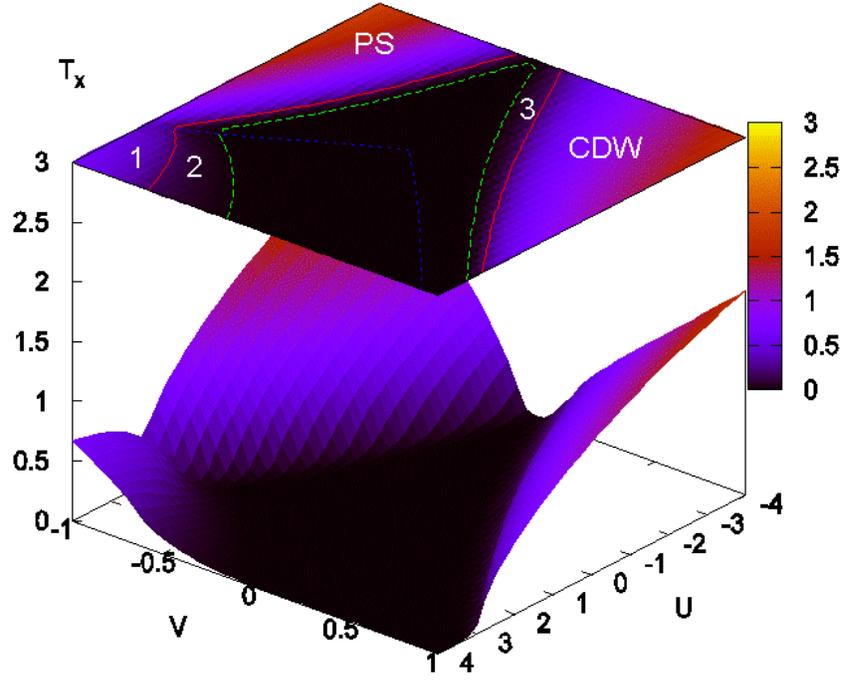}
\end{center}
\caption{The crossover temperature for $n=0.75$. Results for negative values
of the interaction strength are given for illustrative purposes only.
Pairing instabilities may dominate in these cases. In the region outside the
red line, commensurate (1) SDW $\left( \protect\pi ,\protect\pi \right) $,
CDW $\left( \protect\pi ,\protect\pi \right) $ and PS $\left( 0,0\right) $
fluctuations dominate. (The regions of interest are similar to those
observed in Fig~7 at $n=1,$ except that SDW become much less important). In
the area surrounded by the red and green lines, the renormalized classical
fluctuations are incommensurate ISDW in (2) and ICDW in (3). The area inside
the green line boundary is paramagnetic. The blue line separates the regions
where the spin (left and bottom) and charge (right and top) components of
the response functions are dominant.}
\end{figure}

\end{widetext}%
Fig.~8 shows the crossover diagram for $n=0.75$. The fluctuations outside
the red lines are commensurate. The commensurate CDW persists away from
half-filling \cite{Ohta:1994} while the SDW fluctuations are greatly reduced
in the $U>0,V>0$ regime. The fluctuations are incommensurate in the narrow
area between the green and red lines. For $U$ and $V$ inside the green
curve, the system is paramagnetic, which means that neither the spin nor
charge response functions can satisfy the above mentioned criteria even at
very low temperature ($T=0.01$). Finally the blue lines separate
the region of parameter space where the spin fluctuations are larger than
charge fluctuations (left bottom) from the region where the reverse is true
(right top). Commensurate fluctuations have the highest crossover
temperature. The next most important fluctuations are incommensurate. They
exist next to commensurate fluctuations but their crossover temperatures are
rather small.


\section{Conclusion and Summary}

In this paper, we extended the non-perturbative TPSC approach to the case
where we add to the standard Hubbard model a nearest-neighbor interaction $%
V. $ We verified the validity of this Extended Two-Particle Self-Consistent
approach (ETPSC) by comparisons with Quantum Monte Carlo calculations. ETPSC
applies up to intermediate coupling and it satisfies conservation laws and
the Mermin-Wagner theorem. It also includes renormalization of the
irreducible vertices by quantum fluctuations (Kanamori-Br\"{u}ckner).

In the original TPSC, the evaluation of the irreducible vertices requires
one pair-correlation function and its functional derivative. These two
quantities are evaluated by imposing the Pauli principle on two sum rules
that lead to self-consistency conditions. In the presence of $V,$ a similar
procedure for the irreducible vertices allows us to determine the needed
pair correlation functions self-consistently \cite{Bahman}, but not the
functional derivatives of these pair correlation functions. One approach is
to neglect the functional derivatives, in the spirit of the Singwi approach 
\cite{Singwi} to the electron gas, as we did earlier \cite{Bahman}. The new
idea of the present paper is to use particle-hole symmetry to evaluate the
missing functional derivatives. When $V$ is finite, we found that using the
resulting estimate of the functional derivatives and dropping one of the sum
rules between spin and charge fluctuations implied by the Pauli principle
gives better agreement with benchmark QMC than that obtained by neglecting
the functional derivatives \cite{Bahman}. Although particle-hole symmetry
strictly applies only to the nearest-neighbor hopping model at half-filling,
one expects that the approximation will not be too bad away from
half-filling. The results for $n=0.8$ on the bottom right panel of Fig.2 are
encouraging in this respect.

For illustrative purposes, we also presented the crossover diagram of the
two-dimensional extended Hubbard model on the square lattice for $n=1$ and $%
n=0.75$ and a range of values of $V$ and $U.$ Incommensurate fluctuations
appear away from half-filling, but usually they do so at temperatures much
lower than commensurate fluctuations. We highlighted the differences with
mean-field calculations \cite{Murakami}. The crossover temperature indicates
when the fluctuations associated with a self-consistently determined wave
vector become large, or more specifically when these fluctuations enter the
renormalized classical regime. Based on earlier arguments \cite{Vilk2}, we
expect that a pseudogap appears in the renormalized classical regime
whenever the wave vector of the fluctuations can scatter electrons from one
side to the other side of the Fermi surface.

The methodology developed in the present paper can be easily applied to
different lattices and higher dimension. It can be extended also to
longer-range interactions. Note also that the functional derivatives
introduce just two independent unknowns, namely $\frac{\delta g_{\uparrow
\downarrow }(1,1)}{\delta n(1)}$ and $\frac{\delta g_{s\sigma }(1,2)}{\delta
n(1)}=\frac{\delta g_{s\sigma }(1,2)}{\delta m(1)}.$ If there were
independent ways of estimating the compressibility and the uniform spin
susceptibility, that may give, with the Pauli principle, new sum-rules that
could be used to estimate these functional derivatives. That question and
that of multiband models are for future research. We
also plan to apply ETPSC to specific compounds, such as the cobaltates,
where charge fluctuations are important.


\section{Acknowledgments}

We thank S. Hassan and B. Kyung for useful discussions. Computations were
performed on the Elix2 Beowulf cluster in Sherbrooke. 
The present work was supported by the Natural
Sciences and Engineering Research Council NSERC (Canada), the Fonds qu\'{e}b%
\'{e}cois de recherche sur la nature et la technologie FQRNT (Qu\'{e}bec),
the Canadian Foundation for Innovation CFI (Canada), the Canadian Institute
for Advanced Research CIAR, and the Tier I Canada Research Chair Program
(A.-M.S.T.).




\begin{thebibliography}{99}
\bibitem{Oles} K. Rosci\'{s}zewski and A. Ol\'{e}s, J. Phys. Condens. Matter 
\textbf{15}, 8363 (2003).

\bibitem{Imada:2005} Kota Hanasaki, and Masatoshi Imada, J. Phys. Soc. Jpn. 
\textbf{74,} 2769 (2005).

\bibitem{Bickers:1989} N.E.Bickers, and D.J. Scalapino, Ann. Phys. (USA), 
\textbf{193, }206 (1989).

\bibitem{Vilk2} Y. M. Vilk, and A. -M. Tremblay, J. Phys. I \textbf{7}, 1309
(1997).

\bibitem{Vilk1} Y. M. Vilk, Liang Chen and A.-M. Tremblay, Phys. Rev. B 
\textbf{49}, 13267 (1994).

\bibitem{Mahan} G.D. Mahan, \textit{Many-Particle Physics}, 3rd edition,
Section 6.4.4 (Kluwer/Plenum, 2000)

\bibitem{Hankevych} V. Hankevych, B. Kyung, and A.-M. S. Tremblay, Phys.
Rev. \textbf{B} 68, 214405 (2003).

\bibitem{Moukouri} S. Moukouri, S. Allen, F. Lemay, B. Kyung, D. Poulin, Y.
M. Vilk, and A.-M. S. Tremblay, Phys. Rev. B 61, 7887 (2000).

\bibitem{AllenKyung} B. Kyung, S. Allen, A.-M. S. Tremblay, Phys. Rev. B 
\textbf{64}, 075116 (2001).

\bibitem{Zhang} Y. Zhang and J. Callaway, Phys. Rev. B \textbf{39}, 9397
(1989).

\bibitem{Bahman} B. Davoudi and A.M. Tremblay, Phys. Rev. B \textbf{74},
035113 (2006).

\bibitem{Pietig:1999} R. Pietig, R. Bulla, and S. Blawid, Phys. Rev. Lett. 
\textbf{82, }4046 (1999).

\bibitem{Calandra:2002} M. Calandra, J. Merino, and R. H. Mckenzie, Phys.
Rev. B. \textbf{66} 195102 (2002).

\bibitem{Vojta:1999} M. Vojta, R.E. Hetzel, R.M. Noack, Phys. Rev. B \textbf{%
60}, R8417 (1999).

\bibitem{Vojta:2001} M. Vojta, A. Hubsch, R.M. Noack, Phys. Rev. B \textbf{63%
}, 045105 (2001).

\bibitem{Seo:1998} H. Seo and H. Fukuyama, J. Phys. Soc. Japan, \textbf{67},
2602 (1998).

\bibitem{Hoang:2002} A. Hoang, P. Thalmeier, J. Phys.: Condens. Matter 
\textbf{14,} 6639 (2002).

\bibitem{Hellberg:2001} C.S. Hellberg, J. Appl. Phys. \textbf{89}, 6627
(2001).

\bibitem{Fourcade} B. Fourcade and G. Sproken, Phys. Rev. B \textbf{29},
5096 (1984); J. E. Hirsch, Phys Rev Lett. \textbf{53}, 2327 (1984); H. Q.
Lin and J. E. Hirsch, Phys. Rev. B \textbf{33}, 8155, (1986).

\bibitem{Bosch:1988} L. M. del Bosch and L. M. Falicov, Phys. Rev. B \textbf{%
37}, 6037 (1988).

\bibitem{Aichhorn:2004} M. Aichhorn, H.G. Evertz, W. von der Linden, and M.
Potthoff, Phys. Rev. B \textbf{70,} 235107 (2004).

\bibitem{Barktowiak:1995} B M. Bartkowiak, J. A. Henderson, J. Oitmaa, P. E.
de Brito, Phys. Rev. B \textbf{51}, 14077 (1995).

\bibitem{VanDongen:1994} P. G. J. van Dongen, Phys. Rev. B \textbf{49}, 7904
(1994); P. G. J. van Dongen, Phys. Rev. B \textbf{50}, 14016 (1994).

\bibitem{Avella:2004} A. Avella, F. Mancini, Euro. Phys. J. B \textbf{41},
149 (2004).

\bibitem{Avella:2006} A. Avella and F. Mancini, J. Phys. Chem. Sol. \textbf{%
67}, 142 (2006).

\bibitem{Vojta:2002} M. Vojta, Phys. Rev. B \textbf{66}, 104505 (2002).

\bibitem{Kuroki:2006} K. Kuroki, J. Phys. Soc. Japan \textbf{75}, 114716
(2006).

\bibitem{Seo:2006} H. Seo, K. Tsutsui, M. Ogata, and J. Merino, J. Phys.
Soc. Japan \textbf{75, }117707 (2006).

\bibitem{Onari} S. Onari, R. Arita, K. Kuroki, and H. Aoki, Phys. Rev. B 
\textbf{70}, 094523 (2004).

\bibitem{Kishine:1995} J. Kishine, H. Namaizawa, Prog. Theor. Phys. \textbf{%
93}, 519 (1995).

\bibitem{Callaway:1990} J. Callaway, D.P. Chen, D.G. Kanhere, and Qiming Li,
Phys. Rev. B \textbf{42}, 465 (1990) and Physica B \textbf{163}, 127 (1990).

\bibitem{Scalapino} D. J. Scalapino, E. Loh, Rr. and J. E. Hirsch, Phys.
Rev. B \textbf{35}, 6694 (1987).

\bibitem{Esirgen} G. Esirgen and N. E. Bickers, Phys. rev. B \textbf{55},
2122 (1997); \textit{ibid} \textbf{57}, 5376 (1997)

\bibitem{Yan:2006} X.Z. Yan, Phys. Rev. B \textbf{73}, 052501 (2006).

\bibitem{Murakami} M. Murakami, Journal of the Physical Society of Japan 
\textbf{69}, 1113 (2000).

\bibitem{Ohta:1994} Y. Ohta, K. Tsutsui, W. Koshibae, and and S. Maekawa,
Phys. Rev. B \textbf{50}, 13594 (1994).

\bibitem{Yan:1993} X.Z. Yan Phys. Rev. B \textbf{48}, 7140 (1993).

\bibitem{Bari:1971} R.A. Bari, Phys. Rev. B \textbf{3, }2662 (1971).

\bibitem{Micnas} A.M. Ol\'{e}s, R. Micnas, S. Robaszkiewicz, K.A. Chao,
Phys. Lett. A, \textbf{102}, 323 (1984).

\bibitem{Allen:2003} S. Allen, A.-M. Tremblay and Y. M. Vilk, in \textit{%
Theoretical Methods for Strongly Correlated Electrons}, edited by D. S\'{e}n%
\'{e}chal, C. Bourbonnais and A.-M. Tremblay, CRM Series in Mathematical
Physics, Springer (2003);

\bibitem{BaymKadanoff:1962} L. P. Kadanoff and G. Baym, \textit{Quantum
Statistical Mechanics} (Benjamin, Menlo Park, 1962).

\bibitem{Singwi} K. S. Singwi, M. P. Tosi, R. H. Land and A. Sj\"{o}lander,
Phy. Rev. \textbf{176}, 589 (1968).

\bibitem{BickersWhite:1991} N.E. Bickers and S.R. White, Phys. Rev. B 
\textbf{43} 8044 (1991).

\bibitem{AllenTremblay} S. Allen, et A.-M. S. Tremblay, Phys. Rev. B \textbf{%
64}, 075115 (2001).
\end{thebibliography}
\end{document}